\documentclass[pre,twocolumn,showpacs,amsmath,amssymb]{revtex4}

\usepackage{graphicx}% Include figure files
\usepackage{bm}% bold math
\usepackage{setspace,ulem,bm}
\begin{document}

\title{Compaction and dilation rate dependence of stresses
in gas-fluidized beds}
\author{Sung Joon Moon, I. G. Kevrekidis, and
S. Sundaresan\footnote{Corresponding author:
sundar@princeton.edu; 609-258-4583 (tel); 609-258-0211 (fax).}}
\affiliation{Department of Chemical Engineering \&
Program in Applied and Computational Mathematics\\
Princeton University, Princeton, NJ 08544}

%\date{\today}% It is always \today, today,
             %  but any date may be explicitly specified

\begin{abstract}
A particle dynamics-based hybrid model, consisting of monodisperse
spherical solid particles and volume-averaged gas hydrodynamics,
is used to study traveling planar waves (one-dimensional traveling
waves) of voids formed in gas-fluidized beds of narrow cross
sectional areas.
Through ensemble-averaging in a co-traveling frame,
we compute solid phase continuum variables (local volume fraction,
average velocity, stress tensor, and granular temperature)
across the waves, and examine the relations among them.
We probe the consistency between such computationally obtained
relations and constitutive models in the kinetic theory
for granular materials which are widely used in the two-fluid
modeling approach to fluidized beds.
We demonstrate that solid phase continuum variables exhibit
appreciable ``path dependence'', which is not captured by the
commonly used kinetic theory-based models.
We show that this path dependence is associated with the
large rates of dilation and compaction that occur in the wave.
We also examine the relations among solid phase continuum variables
in beds of cohesive particles, which yield the same path dependence.
Our results both for beds of cohesive and non-cohesive particles
suggest that path-dependent constitutive models need to be developed.
\end{abstract}

\maketitle

\section{Introduction}

Flows involving gas-particle mixtures are ubiquitous in
nature and in engineering practice~\cite{gidaspow94,crowe97,
fan98,jackson00}.
Homogeneous flows are often unstable, and spatiotemporal
structures such as traveling waves, clusters and streamers of
particles, and bubble-like voids are commonly observed
in fluidized beds. The origin of the instabilities leading to
these structures has been studied extensively in the
literature~\cite{savage92,gidaspow94,fan98,jackson00,sundar03}.

The number of particles present in these flows is large,
rendering detailed description of the motion of all the
particles impractical.
Hence, macroscopic flow characteristics are often probed
through analysis of continuum models derived by averaging the
equations governing the motion of the individual particles and
the interstitial fluid.
In this approach, the fluid and solid phases are treated as
inter-penetrating continua, and locally-averaged quantities
such as the volume fractions and velocities of fluid and solid
phases appear as dependent variables.
The details of flow at the level of individual particles appear
in the averaged equations through the effective fluid and particle
phase stresses and the interphase interaction force, for which
one must postulate constitutive models.
It is now known that inhomogeneous structures such as traveling
waves and bubble-like voids in fluidized beds can be reproduced
in a qualitatively correct manner through very simple constitutive
models for the above-mentioned quantities
(see e.g., Glasser et al.~\cite{glasser96}).
Quantitative predictions are still elusive in many cases.  

The derivation of constitutive models for these terms has been
studied extensively in the literature. For example, the kinetic
theory of granular materials (KTGM) for
the quantification of the stresses in a rapidly shearing assembly
of particles undergoing only binary collisions has been discussed
by many authors~\cite{savage83,lun84,jenkins85,campbell90,gidaspow94,
koch99,jackson00},
and it has found widespread applications in many gas-particle flow
problems~\cite{gidaspow94,agrawal01}.
Many of the early studies of the kinetic theory focused on spherical,
uniformly sized, non-cohesive particles, while more recently there
have been attempts to extend it to slightly cohesive systems
(see e.g., Kim and Arastoopour~\cite{kim02}).
Particle dynamics simulations of shear flow of nearly homogeneous
assemblies of non-cohesive particles have been used to validate the
kinetic theory~\cite{campbell90,hopkins90,clelland02,curtis04}. 

As mentioned above, gas-particle flows in fluidized beds readily
form inhomogeneous structures, where the particle assembly undergoes
%compaction and dilation alternately in a quasi-periodic manner while
compaction and dilation in an alternating periodic manner while
also undergoing a shear. For example, in fluidized beds, this results
from the flow patterns created by a succession of bubble-like voids
rising through the bed.
Quite frequently, such effects occur over length scales in the range
of 10 - 50 particle diameters~\cite{sundar03,agrawal01}, and the
effect of the rate of compaction or dilation accompanying inhomogeneous
flows at such length scales on the rheological characteristics is
largely unknown. 

While the KTGM does include a bulk viscosity term to account for
the effect of compaction/dilation rate on the solid phase pressure,
the adequacy of this correction remains untested.
A good understanding of the typical rates of compaction and dilation
that can occur in fluidized gas-particle suspensions and their
influence on the stresses is a necessary step in the quest for
constitutive models that can lead to quantitative predictions of
their flow behavior.
The first goal of this study is to analyze a simple, idealized
gas-particle flow problem where a particle assembly undergoes
compaction and dilation in an alternating manner and also manifests
spatial inhomogeneity that is typical of fluidized suspensions,
and probe how the stresses are influenced by compaction or dilation
events.
To this end, we have carried out particle dynamics-based hybrid
simulations of a periodic assembly of uniformly sized, spherical,
non-cohesive, frictional and inelastic particles fluidized by a
gas, and created an inhomogeneous flow pattern that takes the form
of a one-dimensional (vertically) traveling wave (1D-TW).
In this approach, solid phase stresses are determined through
particle-particle interactions, rather than by postulated
constitutive models, and various inter-particle forces such as
cohesive forces can readily be accounted for without introducing
additional assumptions.
By traveling along this wave and averaging over many realizations,
we have extracted the spatial variation of various ensemble-averaged
quantities.
{\it It will be shown that the stresses manifest strong path
dependence.}
When the assembly undergoes dilation, the stress levels are very
small and are essentially independent of the rate of dilation;
in contrast, when compaction occurs, {\it the stresses depend
strongly on the rate of compaction.}

We have also extended these simulations to include inter-particle
attractive forces; specifically, we considered a simple model to
capture van der Waals force between the particles.
It will be shown through an analysis of traveling waves in modestly
cohesive systems that the path dependence is even greater in
cohesive systems. 

The rest of the paper is organized as follows.
The model we use, a particle dynamics model coupled with
volume-averaged gas hydrodynamics, is described in
Sec.~\ref{method}.
The traveling wave structures and properties obtained in
our simulations of beds of non-cohesive particles are
presented in Sec.~\ref{1DTW}.
In Sec.~\ref{path}, solid phase continuum variables across
the waves are analyzed and assumptions used in KTGM are probed.
The waves formed in beds of moderately cohesive particles
are analyzed in Sec.~\ref{cohesion}, followed by conclusions
in Sec.~\ref{conclusion}. 

\section{Method}
 \label{method}

\subsection{DEM-based hybrid model}

The solid phase is modeled as a collection of discrete particles
of ``soft'' monodisperse spheres (the deformation is accounted for
by overlaps), whose individual trajectories are computed by
integrating classical equations of motion.
Such an approach is often referred to as the soft-sphere molecular
dynamics simulation or the discrete element method (DEM).
%we apply this strategy to macroscopic particles.
When a pair of objects (particles or system boundaries) get into
contact with each other, the interaction force is analyzed in the
normal and tangential directions separately, $\mathbf{F}_{cont} =
(\mathbf{F}_n,\mathbf{F}_s)$. We adopt the so-called spring-dashpot
model of Cundall and Strack~\cite{cundall79}, using a Hookean spring
\begin{eqnarray}
 \label{normal}
{\mathbf F}_n & = & (k_n \Delta_n - \gamma_n v_n) \hat{\mathbf n}, \\
{\mathbf F}_s & = & -{\rm sign}(v_s)\times{\rm min(}k_t \Delta_s, \mu |{\mathbf F}_n|) \hat{\mathbf s},
\end{eqnarray}
where $k_n$ is the spring stiffness in the normal direction 
($\hat{\mathbf n}$ is the corresponding unit vector,
pointing from the contact point toward the particle center);
$\Delta_n$ is the overlap of the particles in the normal direction;
$\gamma_n$ is the damping coefficient determined by $k_n$ and
the normal coefficient of restitution $e$ characterizing the
inelastic kinetic energy loss upon contact; and $v_n$ is the
normal component of the relative velocity at contact.
The interaction in the tangential direction is modeled by a
spring and a slider, where the tangential stiffness $k_t$ is
determined by $k_n$ and the Poisson's ratio of the material $\nu_P$
[a value of 0.3 is used for all the computations in our study;
$k_t = 2k_n(1-\nu_P)/(2-\nu_P)$];
$\Delta_s$ is the tangential displacement from the
initial contact; $v_s = {\mathbf v}_s\cdot\hat{\mathbf s}$
is the tangential component of the relative velocity at contact
[${\mathbf v}_s
= \hat{\mathbf n}\times({\mathbf v}_{ij}\times\hat{\mathbf n})$];
and $\hat{\mathbf s}$ is the unit vector in the tangent plane
collinear with the component of the relative velocity at contact.
The total tangential force is limited by the Coulomb frictional
force $\mu |\mathbf{F}_n|$, where $\mu$ is the coefficient of
friction.
More details on the model equations can be found
elsewhere~\cite{moon05c}.

The equation of motion for each particle accounts for the
gas phase effects following the volume-averaged
hydrodynamics~\cite{tsuji93,hoomans96,xu97}:
\begin{equation}
 \label{eachgrain}
m_p{d\mathbf{v}_p \over dt} = m_p\mathbf{g} + \mathbf{F}_{cont} +
\mathbf{F}_c + {V_p \over \phi}\beta(\phi)\left(\mathbf{u}_g -
\mathbf{v}_p \right)- V_p \nabla p,
\end{equation}
where $m_p$ and $\mathbf{v}_p$ are individual particle mass
and velocity, respectively.
The right-hand side includes various forces acting on the
particle, the last two terms arising from the gas-solid
two-way coupling;
the total force acting on the particles due to the fluid is
commonly partitioned into the local drag part and the effective
buoyant part, as was done here (see e.g., Ye et al.~\cite{ye04}).
The first term is the body force due to gravity, where
$\mathbf{g}$ is the gravitational acceleration, and the second
term is aforementioned contact force.
The third term accounts for the van der Waals cohesive force
$\mathbf{F}_c \approx - {A \over 12}{r \over s^2}\hat{\mathbf{n}}$,
following Hamaker's formula~\cite{israel97} in the limit of
$s \ll r_1 = r_2$ (monodisperse spheres),
where $A$ is the Hamaker constant,
$s$ is the inter-surface distance, and $r_1$ and $r_2$ are 
the (identical) radii of particles undergoing interaction.
We avoid the singularity of this formula at contact by
introducing a minimum separation distance $\delta^*$ of
0.4 nm~\cite{seville00},
below which the same attractive force is applied; i.e.,
$\mathbf{F}_c = {\rm max}[-{A \over 12}{r \over s^2},
- {A \over 12}{r \over (\delta^*)^2}]\hat{\mathbf{n}}$.
The fourth term accounts for the drag force,
where $\beta$ is the interphase momentum transfer coefficient,
$\phi$ is the local particle phase volume fraction,
$\mathbf{u}_g$ is the local-average gas phase velocity, and
$V_p$ is the individual particle volume.
The last term accounts for the hydrodynamic force due to the
gradually varying part of the pressure field, where $p$ is the
local-average gas phase pressure.

In general, the gas phase variables in the last two terms
are obtained by simultaneously integrating the balance equations
for gas-solid mixtures;
however, for the case of 1D incompressible gas phase,
Eq.~(\ref{eachgrain}) can be significantly reduced by
considering the 1D continuity relation:
\begin{equation}
 \label{1Dcont}
(1 - \phi) \mathbf{u}_g + \phi \mathbf{u}_s = \mathbf{U}_s,
\end{equation}
and the reduced momentum balance equation:
\begin{equation}
 \label{simpleNS}
0 =  -(1 - \phi)\nabla p + \beta(\phi) \left(\mathbf{u}_s - \mathbf{u}_g\right),
\end{equation}
where $\mathbf{u}_s$ is the coarse-grained particle velocity,
and $\mathbf{U}_s$ is the superficial gas flow velocity.
After some manipulation [using Eqs.~(\ref{1Dcont}) and (\ref{simpleNS})],
Eq.~(\ref{eachgrain}) can be reduced to the following one,
where gas phase effects appear as additional terms in the
individual equation of motion, involving solid phase coarse-grained,
continuum variables~\cite{moon05c}:
\begin{eqnarray}
\label{oneDeq}
m_p{d\mathbf{v}_p \over dt} = && m_p\mathbf{g} + \mathbf{F}_{cont} + \mathbf{F}_c + {V_p \over \phi}\beta(\phi) \times \nonumber \\
&& \left[(\mathbf{u}_s - \mathbf{v}_p) - {1\over (1 - \phi)^2}(\mathbf{u}_s - \mathbf{U}_s) \right].
\end{eqnarray}

For the momentum transfer coefficient $\beta$, we use an empirical
expression proposed by Wen and Yu~\cite{wen66}:
\begin{equation}
 \label{beta}
\beta = {3\over 4} C_D {\rho_g \phi(1 - \phi) |\mathbf{u}_g - \mathbf{u}_s| \over d_p} (1 - \phi)^{-2.65},
\end{equation}
where $C_D$ is the drag coefficient, $\rho_g$ is the gas phase
mass density, and $d_p$ is the particle diameter.
The drag coefficient proposed by Rowe~\cite{rowe61} is employed
\begin{equation}
C_D = \left\{ \begin{array}{ll}
{24 \over Re_g}\left(1 + 0.15Re_g^{0.687}\right), & Re_g < 1000, \\
0.44, & Re_g \geq 1000,
\end{array}
      \right.
\end{equation}
where 
\begin{equation}
Re_g = {(1 - \phi)\rho_g d_p |\mathbf{u}_g - \mathbf{u}_s|
\over \mu_g},
\end{equation}
and $\mu_g$ is the gas phase viscosity.
We further simplify these formulas using the assumption of
$Re_g \ll 1$, as we consider only small particles.

\subsection{Fully periodic box}
 \label{periodic}

Traveling waves in a bed of a finite depth (as in experimental
systems) are hardly perfectly periodic.
In order to eliminate such an imperfection from the wave,
we consider an idealized geometry of a fully periodic (in all
three directions) small box of height $L$ which is commensurate
with the wavelength $\lambda$ in a deep fluidized bed~\cite{moon05c}.
The periodic boundary conditions for both lateral directions
are achieved with the usual geometrical wrapping.
In the direction of gravity, we adopt the following procedure:
For the fully fluidized states that we are interested in, the total
weight of the bed is supported by the pressure drop, which
can be written as
\begin{equation}
 \label{3dperiodic}
p|_{z=0}-p|_{z=L} = \rho_s g \phi_{avg} L,
\end{equation}
where $\rho_s$ is the mass density of the solid phase,
$g = |\mathbf{g}|$ is the acceleration due to gravity,
and $\phi_{avg}$ is the average volume fraction of the bed.
Solving Eq.~(\ref{3dperiodic}) together with the 1D continuity
relation Eq.~(\ref{1Dcont}) and the momentum balance equation
Eq.~(\ref{simpleNS}) yields an appropriate value for the superficial
gas flow velocity $\mathbf{U}_s$ that fully fluidizes the bed
in this small periodic box.
We use this value of $\mathbf{U}_s$ in Eq.~(\ref{oneDeq}).
%in order to compute a fully fluidized state in a fully periodic box.

\subsection{Nondimensionalization}
 \label{nondim}

Casting Eqs.~(\ref{oneDeq}) and (\ref{beta}) in a dimensionless form,
using $\rho_s$, $d_p$, $\sqrt{gd_p}$, $\sqrt{d_p/g}$ as characteristic
density, length, velocity, and time, one obtains the following
nondimensional groups (arrows indicate changes in the notation
from dimensional variables to nondimensional variables that will be
used henceforth):
\begin{eqnarray*}
k_n &\leftarrow& \frac{k_n}{\rho_s g d_p^2}, ~~~~~~~~{\rm spring~stiffness}\cr
U_s &\leftarrow& \frac{U_s}{\sqrt{g d_p}}, ~~~~~~~~{\rm superficial~gas~flow~rate}\cr
\delta &\equiv& \frac{\delta^*}{d_p}, ~~~~~~~~~~~~~{\rm scaled~minimum~separation~distance}\cr
Bo &\equiv& \frac{A}{4\pi \rho_s g d_p^2\delta^2}, ~~~{\rm cohesive~Bond~number}\cr
St &\equiv& \frac{\rho_s g^{1/2} d_p^{3/2}}{\mu_g}, ~~~{\rm Stokes~number}
\end{eqnarray*}
together with nondimensional parameters, namely $e$, $\mu$, $\nu_P$,
and $L$ the nondimensional bed height.
We characterize particles by usual values ($e = 0.9; \mu = 0.1;
\nu_P = 0.3$) in experiments,
except when we consider ``ideal'' particles in Figs.~\ref{ideal}
and \ref{shear_ideal}.
In the following sections, all the results are presented in
nondimensional form.

\begin{figure}[t]
\begin{center}
\includegraphics[width=.77\columnwidth]{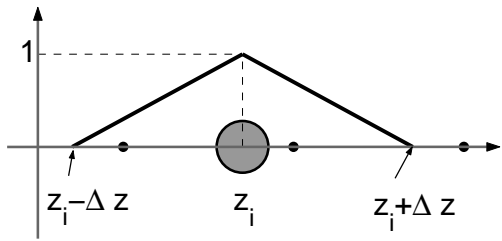}
\end{center}
\caption{\label{halo}
A schematic of a halo function in Eq.~(\ref{halo_func}), applied to
a particle (a large circle) whose center is located at $z = z_i$,
which is used to compute solid phase coarse-grained continuum variables.
Three dots along the abscissa represent nearby grid points.
The halo function linearly decreases from the unity at
$z = z_i$ to zero at $z = z_i \pm \Delta z$.
}
\end{figure}

\subsection{Computation of solid phase continuum variables}

Solid phase coarse-grained continuum variables at 1D discrete
grid points separated by $\Delta z$ are computed by distributing
the particle mass and momenta values to two nearby grid points
using a halo function $h$ that linearly decreases to zero
around the $i$th particle located at $z = z_i$:
\begin{equation}
 \label{halo_func}
h(z;z_i) = \left\{ \begin{array}{ll}
                1-{|z-z_i|/ \Delta z} &~~~ {\rm for~} |z-z_i| < \Delta z,\\
                0 &~~~ {\rm otherwise,}
                 \end{array}
        \right.
\end{equation}
where
$\Delta z$ is the grid spacing.
It is readily seen that $h$ has the property that the solid
phase quantities of each particle are distributed to the two
nearby grid points, inversely proportional to the distance to
each grid point.
The number density $n [= (6/\pi)\phi/d_p^3]$ and $\mathbf{u}_s$,
on the grid point $z_0$ are then defined simply as
\begin{eqnarray}
n(z_0) & = & \sum_{i= 1}^N h(z_0;z_i),\\
n(z_0){\mathbf u}_s(z_0) & = & \sum_{i= 1}^N h(z_0;z_i){\mathbf v}_{p,i},
\end{eqnarray}
where $z_i$ is the $i$th particle location.
Following the same procedure, it is straightforward to compute
the granular temperature tensor \uuline{$T$}:
\begin{equation}
\uuline{T} = \left<({\mathbf v}_{p,i}-{\mathbf u}_s)\otimes
({\mathbf v}_{p,i}-{\mathbf u}_s) \right>,
\end{equation}
or the scalar granular temperature $T$ [$\equiv {1\over D}{\rm Tr}(\uuline{T})$],
where $D$ is the dimensionality, $\otimes$ is the dyadic tensor
product, and ``Tr'' denotes the trace (of a tensor).

The full stress tensor consists of a kinetic or dynamic part and
a virial or static part~\cite{latzel00}:
\begin{equation}
\uuline{\sigma} = {1 \over V} \left[\sum_i m_i \mathbf{\widetilde{v}}_{p,i}
\otimes \mathbf{\widetilde{v}}_{p,i} - \sum_{c \in V} \mathbf{f}_c
\otimes \mathbf{l}_c \right ],
\end{equation}
where $\mathbf{\widetilde{v}}_{p,i} = \mathbf{v}_{p,i} - \mathbf{u}_s$
%where $\mathbf{\widetilde{v}}_{p,i} = \mathbf{v}_{p,i} - <\mathbf{v}_{p,i}>$
is the fluctuating velocity,
$\mathbf{f}_c$ is the force between contacting particles 1 and 2,
and $\mathbf{l}_c = \mathbf{r}_{1} - \mathbf{r}_{2}$ is the displacement
vector between the centers of the particles under consideration.
The second term is summed over all the contacts in the averaging
volume $V$.

Solid phase continuum variables are computed, using the above halo
function, on the nodes separated by the grid spacing $\Delta z$.
In order to compute smoothly varying continuum variables across
a bed, this procedure is repeated on many subgrid points in
between the grid points, by translating all the nodes by a small
amount.
As a result, all the continuum variables are computed on uniformly
distributed points separated by $\Delta z/10$ throughout the bed.

\begin{figure}[t]
\begin{center}
\includegraphics[width=.77\columnwidth]{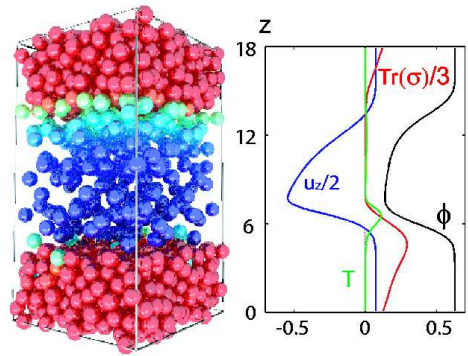}
\end{center}
\caption{\label{raytraced}
(Color online)
Left panel:
A snapshot of non-cohesive particles ($Bo = 0$) in a fully
periodic (in all three directions) box ($10\times10\times18$)
whose height is commensurate with one wavelength of a
one-dimensional traveling wave (left panel).
Right panel:
Corresponding coarse-grained solid phase continuum variables:
local volume fraction $\phi$, vertical component of the locally
averaged velocity $u_z$, granular temperature $T$, and the average
normal stress ${\rm Tr} (\uuline{\sigma})/D$, where $D = 3$ is
the dimension.
[All quantities are dimensionless.]
}
\end{figure}

\section{Simulation results}
 \label{results}

We consider gas-fluidized beds of non-cohesive as well as
cohesive particles in fully periodic boxes of a narrow
square-shaped $10 \times 10$ cross sectional area.
% $10d_p \times 10d_p$.
The height of the box $L$, which is the same as the wavelength
$\lambda$ in a deep fluidized bed, is set to be $18$ in most
cases; it is varied when we study the effect of the wavelength
in Fig.~\ref{lambda}.
We also set $St$ at 55 in most cases, except when $St$
is varied in Fig.~\ref{stokes}.
Grid spacing for continuum variables $\Delta z$ is set to be
$1.5$ in all the computations we will present.
We have checked that the quantitative results slightly vary
when a different value is used for $\Delta z$, but the main
results remain the same.
We also have confirmed that the results are virtually
independent of the cross sectional area~\cite{moon05c}.

We obtain smooth wave profiles by transforming the bed to
the co-traveling frame (with the wave), and averaging continuum
variables over hundreds of snapshots of fully-developed waves
at different instances (as well as spatial smoothing through
the halo function).
%A comparison with particular experimental data can be made
%by proper conversions of variables at given parameters values.
When beds of cohesive particles are considered (Sec.~\ref{cohesion}),
we characterize the level of cohesion by the cohesive Bond number $Bo$,
which is defined to be the ratio of the maximum cohesive force
(at contact) to the gravitational force acting on the particle,
namely the weight of the particle.

\begin{figure}
\begin{center}
\includegraphics[width=.7\columnwidth]{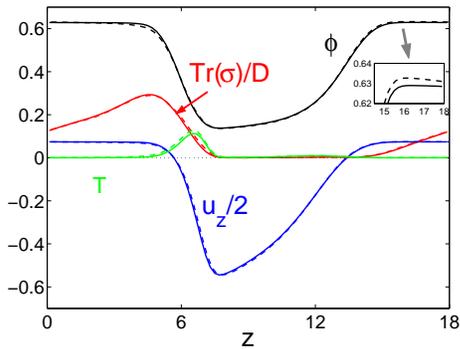}
\end{center}
\caption{\label{ideal}
(Color online)
Continuum variable profiles across fully developed waves
formed in beds of ``realistic'' particles
(solid lines: $e = 0.9, \mu = 0.1$) and of ``ideal'' particles
(dashed lines: $e = 1.0, \mu = 0.0$),
where $\phi_{avg}$ is set to be 0.44 for both cases;
there are only slight differences between the two cases.
The wave travels from left to right, and gravity is acting
toward the negative $z$-direction. 
($Bo = 0; St = 55; k_n = 2.0 \times 10^5$.)
[All quantities are dimensionless.]
}
\end{figure}

\subsection{Traveling planar waves}
 \label{1DTW}

When beds are subject to a superficial gas flow velocity above
the minimum fluidization rate ($\mathbf{U}_s > \mathbf{U}_{mf}$),
homogeneously expanded states are often unstable, and beds
form inhomogeneous structures, such as bubbles~\cite{sundar03}.
In beds of narrow cross sectional areas, bubbling occurs in the
form of propagating planar waves (or 1D-TW) of voidage~\cite{duru02},
which are reproduced in our model.
%, arises from this volume fraction variation.
A snapshot of a bed of non-cohesive particles, forming 1D-TW,
illustrates that particles in the upper plug ``rain down'' through
the void region, and accumulate at the top of the lower plug.
As a result, the void region propagates up (Fig.~\ref{raytraced};
red particles are moving up and blue ones are moving down, in
online color figure).
We use the aforementioned halo function (spatial smoothing) and
time averaging, to compute the corresponding continuum variables,
which are shown in the right panel.
For notational convenience, we drop a subscript of $\mathbf{u}_s$
to represent the coarse-grained solid phase velocity by $\mathbf{u}$
(and represent its z-component by $u_z$).

We start by varying the particle dissipation parameters ($e$
and $\mu$) to probe the sensitivity of the wave profiles to them.
Our simulations reveal that the wave profiles hardly change with
the dissipation parameters (Fig.~\ref{ideal}).
As the particles become more frictional, the plateau volume
fraction values in the plug get smaller, since frictional particles
pack more loosely in the lower plug because of frictional
resistance to rotations (see an inset in Fig.~\ref{ideal}).
However, the change is very small, because the inertia of the
particles is much larger than that of the fluid in gas-fluidized
beds (hence particles accumulate with larger velocities,
compared to those in liquid-fluidized beds).
It is expected that the effect would be more pronounced for
smaller particles.
Li and Kuipers~\cite{li03,li05} showed that heterogeneous structures
do exist in ``ideal'' particles ($e = 1.0$, $\mu = 0.0$), and that
dissipative collisions dramatically intensify the heterogeneity
of the structures.
We do not see such dramatically increased effects;
the waveform changes only slightly, which suggests that the
wave mainly arises from the instability associated with the
interplay between solid phase inertia and the nonlinear
volume-fraction-dependent gas drag, as argued by
Jackson~\cite{jackson63}.
Inelastic collisions themselves indeed lead to inhomogeneous
structures through a well-known ``clustering
instability''~\cite{goldhirsh93}; however, our simulation
suggests that this is a secondary effect in the formation of 1D-TW.

\begin{figure}[t]
\begin{center}
\includegraphics[width=.7\columnwidth]{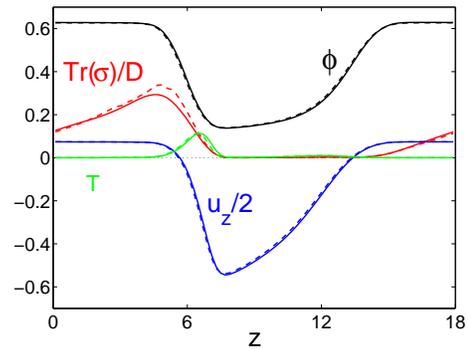}
\end{center}
\caption{\label{stiffness}
(Color online)
Continuum variable profiles across fully developed waves
formed in beds of two types of particles with different
stiffness (solid lines: $k_n = 2.0 \times 10^5$;
dashed lines: $k_n = 2.0 \times 10^8$),
where $\phi_{avg}$ is set to be 0.44 for both cases;
only slight differences between the two cases.
The integration time stepsize for the case of
$k_n = 2.0 \times 10^5$ is larger by a factor of 30.
($e = 0.9; \mu = 0.1; Bo = 0; St = 55; \phi_{avg} = 0.44$.)
[All quantities are dimensionless.]
}
\end{figure}

\begin{figure}[b]
\begin{center}
\includegraphics[width=.7\columnwidth]{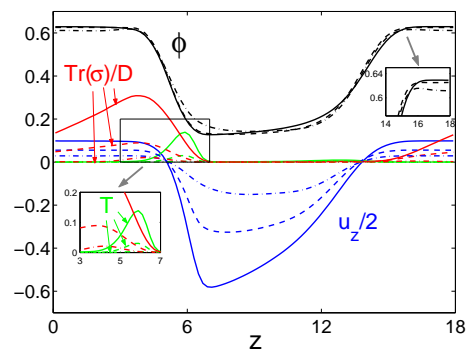}
\end{center}
\caption{\label{stokes}
(Color online)
Continuum variable profiles across fully developed waves
(of the same wavelength), formed in beds of non-cohesive particles
($Bo = 0; e = 0.9; \mu = 0.1; \phi_{avg} = 0.36$) at different
Stokes numbers ($St =$ 55 for solid lines; 28 for dashed lines;
14 for dot-dashed lines).
Volume fraction profiles are nearly the same for the three cases,
while all the other variables decrease in magnitude, as $St$
decreases.
[All quantities are dimensionless.]
}
\end{figure}

\begin{figure*}[t]
\begin{center}

\includegraphics[width=.6\columnwidth]{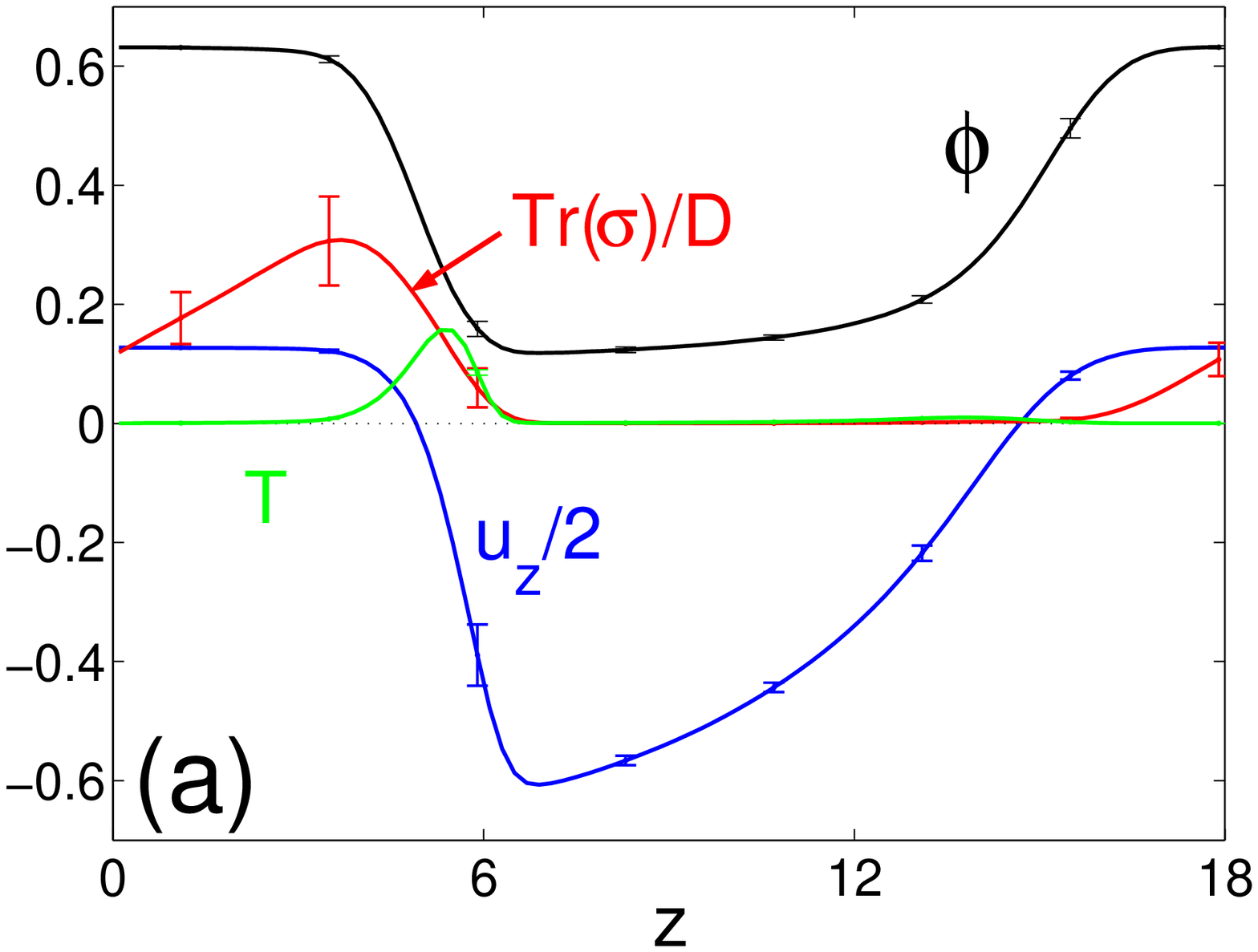}
\includegraphics[width=.6\columnwidth]{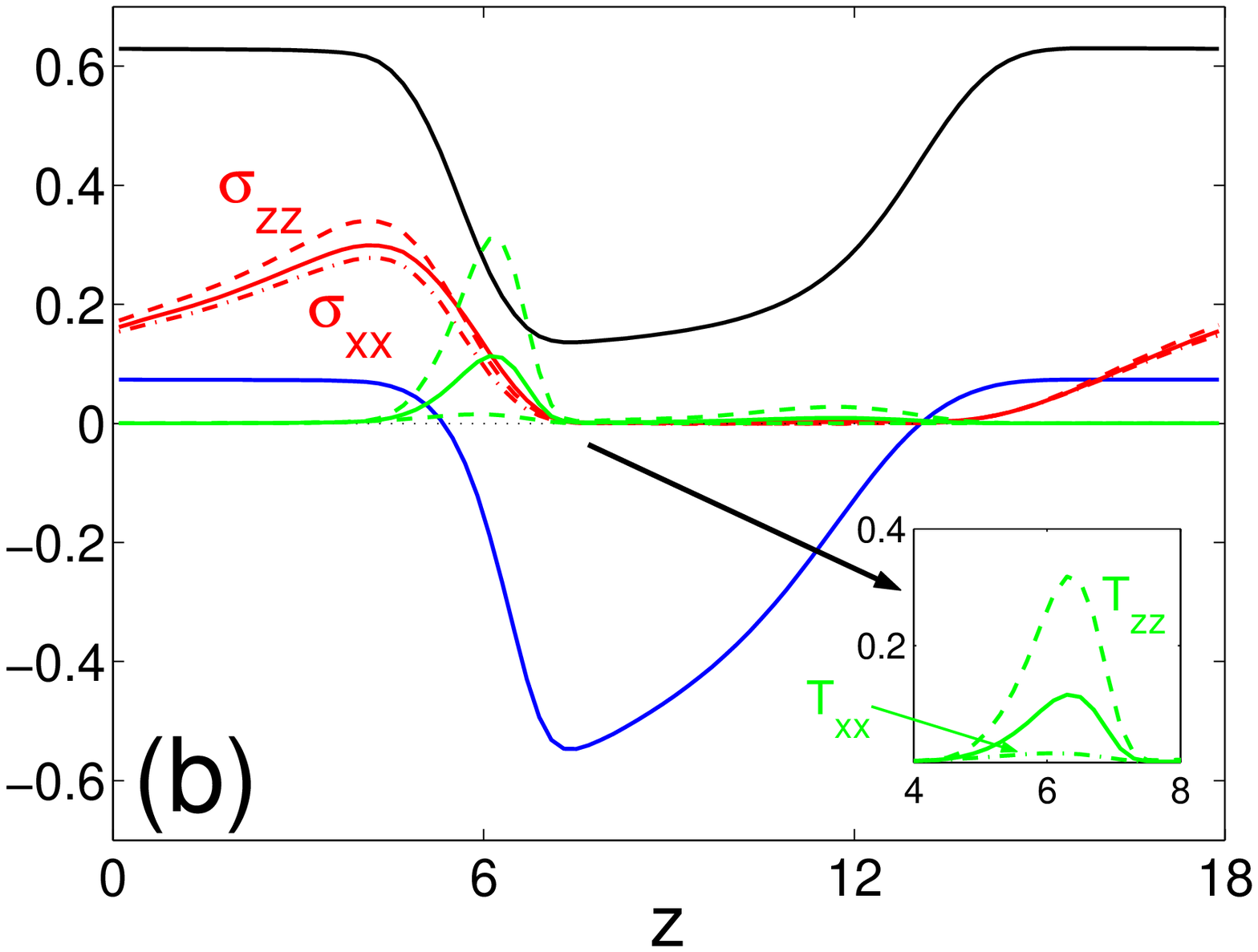}
\includegraphics[width=.6\columnwidth]{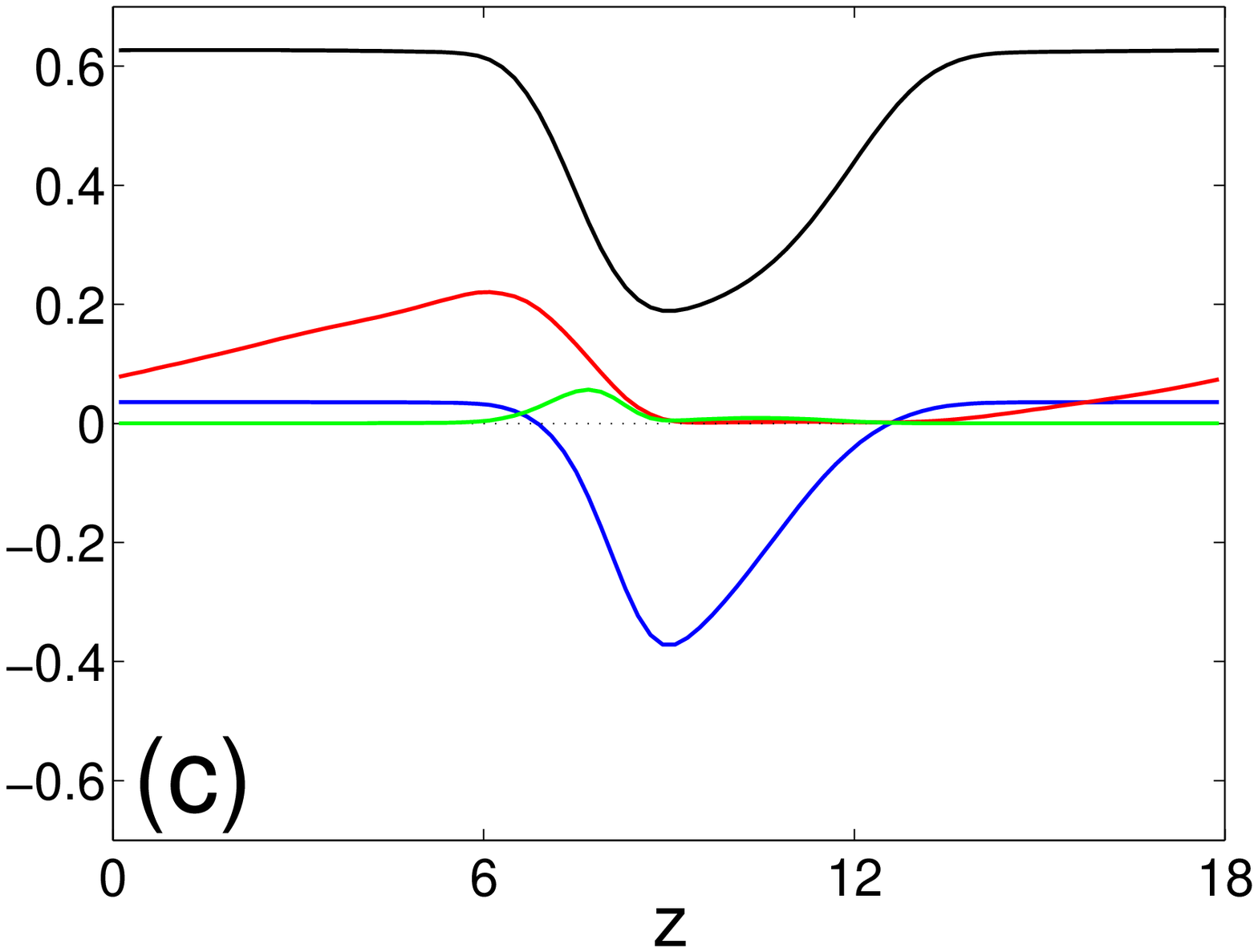}
\end{center}
\caption{\label{vfs}
(Color online)
Profiles of solid phase continuum variables for fully developed
waves formed in beds of non-cohesive particles ($Bo = 0; St = 55;
e = 0.9; \mu = 0.1$)
in a fully periodic box of $10\times10\times18$, shown for three
cases of different average volume fractions
(a) $\phi_{avg} = 0.36$; (b) $\phi_{avg} = 0.44$;
(c) $\phi_{avg} = 0.52$.
Vertical bars in (a) represent the standard deviation of
time-averaged data (over 500 snapshots).
Different diagonal components of the stress tensor $\sigma_{xx}$
(dot-dashed line) and $\sigma_{zz}$ (dashed line) (and those of
the granular temperature tensor in inset) are shown separately
in (b).
[All quantities are dimensionless.]
}
\end{figure*}

One {\it ad hoc} value used in our simulation is the spring
stiffness $k_n$ (as is the case in most other DEM simulations).
In principle, it can be computed from the Young's modulus of
the material under consideration; however, such values limit
the integration step to extremely small sizes,
which leads to expensive computations.
We used two values of $k_n$ differing by three orders of
magnitude (while all the other parameters are kept the same),
and compared the solid phase continuum variables
(Fig.~\ref{stiffness}).
We find that the results are nearly the same, except that the
average normal stress Tr$(\uuline{\sigma})/D$ is somewhat larger
when a stiffer spring constant value is used; however, the
difference in the stress for the two cases is smaller than its
level of fluctuation [which is shown later in Fig.~\ref{vfs} (a)].
For the parameters used in Fig.~\ref{stiffness}, the duration
of head-on collision
$\Delta t_c = \pi/\sqrt{({k_n\over m^*})(1+{1\over \alpha^2})}$,
where $m^*$ is the reduced mas and $\alpha = \pi/\log e$,
corresponds to $8.2\times 10^{-6}$sec and $2.6\times 10^{-7}$sec
for $k_n = 2.0\times 10^5$ and $2.0\times 10^8$, respectively;
it is reasonably small even for the smaller value of $k_n$.
For all the rest of the computations, we will use the smaller $k_n$
value ($\sim 10^5$), as it allows us to use an integration timestep
size ($\sim 10^{-2}\Delta t_c$, which scales with $k_n^{-1/2}$
as shown above) larger by a factor of 30.

\begin{figure}[t]
\begin{center}
\includegraphics[width=.65\columnwidth]{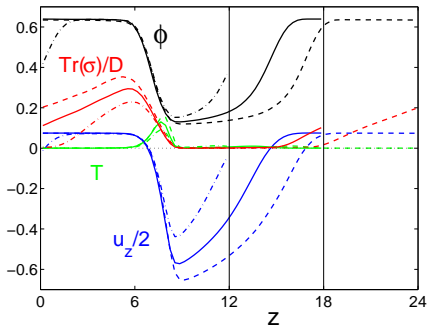}
\end{center}
\caption{\label{lambda}
(Color online)
Continuum variable profiles across fully developed waves of
different wavelength $\lambda$ ($\lambda = 12$ for dot-dashed lines;
$18$ for solid lines; $24$ for dashed lines), formed in beds
of non-cohesive particles ($e = 0.9; \mu = 0.1; Bo = 0; St = 55;
\phi_{avg} = 0.44$).
The plateau volume fractions remain the same, while the amplitudes
of the volume fraction profiles decrease as $\lambda$ decreases.
Vertical thin solid lines at $z = 12$ and 18 represent the system
boundaries for the cases of the same wavelengths.
[All quantities are dimensionless.]
}
\end{figure}

\begin{figure}[t]
\begin{center}
\includegraphics[width=.7\columnwidth]{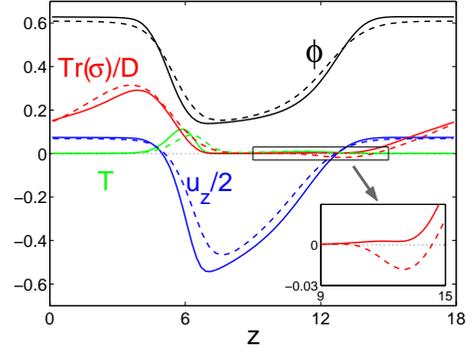}
\end{center}
\caption{\label{bos}
(Color online)
Profiles of solid phase continuum variables obtained for
non-cohesive ($Bo = 0$, solid lines) and slightly cohesive
($Bo = 2$, dashed lines) particles,
with the same average volume fraction, $\phi_{avg} = 0.44$.
The inset shows a blow-up of the region where the stress of
the cohesive bed becomes tensile. ($e = 0.9; \mu = 0.1; St = 55$.)
%The tensile strength measured in the bed is much smaller than
%the estimate of the Rumpf's model (see text).
}
\end{figure}

For the wave profiles shown in Figs.~\ref{ideal} through \ref{lambda},
\ref{dudz}, and \ref{bos}, the gravity is acting leftward ({\it cf.}
right panel in Fig.~\ref{raytraced}).
In sustained traveling waves, particles in the upper plug rain
down through the void region, and the waves travel to right.
In the rest of the paper, we will refer the region where particles
fall off from the bottom of upper plug (characterized by $du_z/dz>0$)
as the {\it dilation} region, and where the particles accumulate
at the top of lower plug ($du_z/dz<0$) as the {\it compaction} region.
Let us track the particle phase continuum variables along the
direction of gravity, starting from the dilation region:
Both the granular temperature and the particle phase stresses
are negligibly small across the void region.
As a stream of particles collide with the lower plug (in the
compaction region), both $T$ and the stress rapidly increase.
The particles quickly become solid-like in the compaction region
through collisional dissipation, and $T$ starts to decrease
(before the stress does).
The collisional stress becomes dominant, and it keeps increasing
until the volume fraction reaches its plateau value.
Inside a plug the stress monotonically decreases, as particles
gradually lose contacts and eventually fall off.

Before analyzing the wave profiles any further, we compare them
at different Stokes numbers by varying the gas phase viscosity
(Fig.~\ref{stokes}).
For a fixed wavelength, the volume fraction profiles change only
slightly; as $St$ decreases (by increasing $\mu_g$ by factors of
2 and 4), the plateau value decreases and the overall profile
gets smoothed a little bit.
More pronounced gas drag makes the magnitude of all other solid
phase quantities decrease by comparable factors.
Quantitative changes in variables (volume fraction, granular
temperature, and average normal stress) are better shown in insets
of Fig.~\ref{stokes}.
We will set $St = 55$ for the rest of the computations, which
corresponds to simulating with usual air.

The volume fraction profiles for a range of average volume fraction
$\phi_{avg}$ [Fig.~\ref{vfs} (a) through (c)] illustrate that the
plateau values in the close-packed regions remain nearly the same,
while both the depth (wave amplitude) and the width of the void region
gradually decrease with increasing $\phi_{avg}$ (when the wavelength
is kept the same).
Both the solid phase average normal stress in the compaction region
(at the top of the lower plug) and the maximum average speed,
decrease as $\phi_{avg}$ increases.
It is worth noting that the solid phase pressure exhibits large
fluctuations, whereas all the other variables ($\phi$,
$\mathbf{u}$, and $T$) exhibit relatively small fluctuations
[Fig.~\ref{vfs} (a)].
In all cases, in the compaction region, the {\it zz}-component of
the stress tensor is larger than the other components
({\it xx-} and {\it yy-}) which are the same due to symmetry
[it is shown only in Fig.~\ref{vfs} (b)].
The observed fluctuations of the stress are quantitatively comparable
with the difference between $\sigma_{zz}$ and $\sigma_{xx}$; however,
the difference is observed consistently in average quantities
for all cases.
The anisotropy is significantly large in the granular temperature
tensor $\uuline{T}$ [see an inset in Fig.~\ref{vfs} (b)], but it
is limited to a narrow region in space.
The first normal stress difference in sheared granular flows is
shown to be a Burnett order effect~\cite{sela98}.
However, the normal stress difference that we observe in a bubbling
fluidized bed has a different origin (see Appendix for details).

\begin{figure}[t]
\begin{center}
\includegraphics[width=.7\columnwidth]{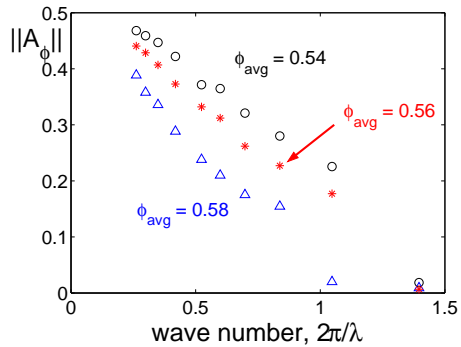}
\end{center}
\caption{\label{amps}
The wave amplitudes (of the volume fraction profiles) obtained
for various wavelengths ($4.5 \leq \lambda \leq 24$) at
three different average volume fractions.
($\phi_{avg}$ = 0.54, 0.56, and 0.58;
$e = 0.9; \mu = 0.1; Bo = 0; St = 55$.)
}
\end{figure}

The continuum variables for waves of different wavelengths, for a
fixed value of average volume fraction, are shown in Fig.~\ref{lambda}.
Once the wavelength is sufficiently long, the same plateau volume
fraction values are obtained for all three cases.
Solid lines in Fig.~\ref{lambda} correspond to the data in
Fig.~\ref{vfs} (b).
From Figs.~\ref{vfs} and \ref{lambda} we see that the plateau
volume fraction values are the same for ranges of average volume
fractions and wavelengths.
This is precisely what Glasser et al.~\cite{glasser96} obtained in
their simulations of continuum models. 
In their analysis, as the wavelength was decreased, the amplitude
of the wave (i.e. of the volume fraction) decreased to zero and
the uniform state became stable at a Hopf bifurcation point.
We simulate the waves for a range of wavelengths, and compute
amplitudes of volume fraction profiles (Fig.~\ref{amps}).
Our analysis yields similar results to those which were obtained
in the continuum model, but the amplitudes at small wavelengths are
finite and the bifurcations are imperfect because of the discrete
nature of the model.
No further resolution of the data at larger wave numbers is
available, as the wavelength increases discretely by
$\Delta z (= 1.5)$ (i.e. $\lambda = 4.5$ and $6.0$ were used).

\subsection{Particle phase stresses and viscosities,
and path dependence}
 \label{path}

We use the data in Fig.~\ref{vfs} to computationally obtain
constitutive relations, and assess consistency with
the assumptions used in the KTGM, which is widely used for
two-fluid models~\cite{gidaspow94,jackson00}.

\begin{figure*}[t]
\begin{center}
\includegraphics[width=.6\columnwidth]{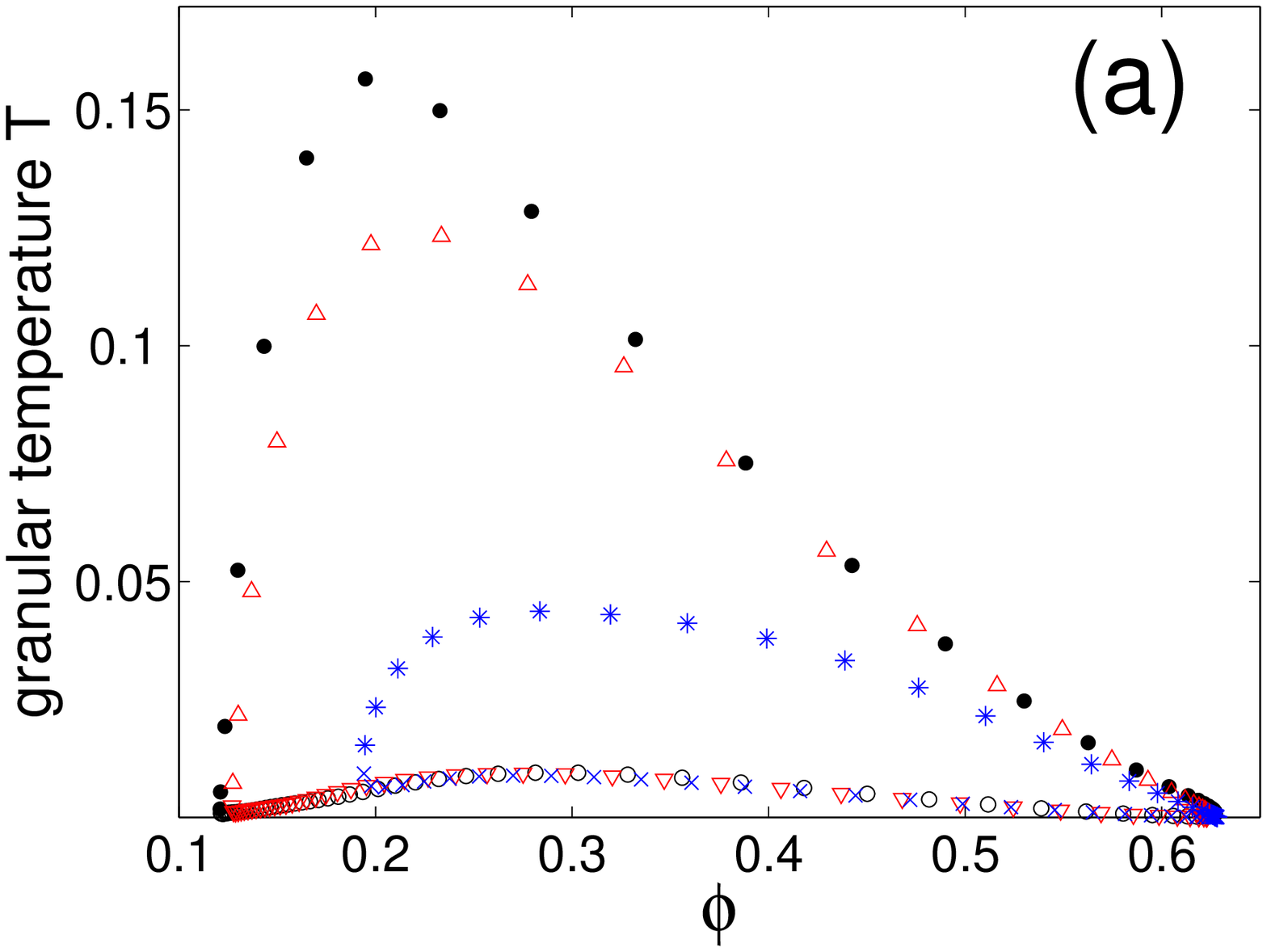}
\includegraphics[width=.6\columnwidth]{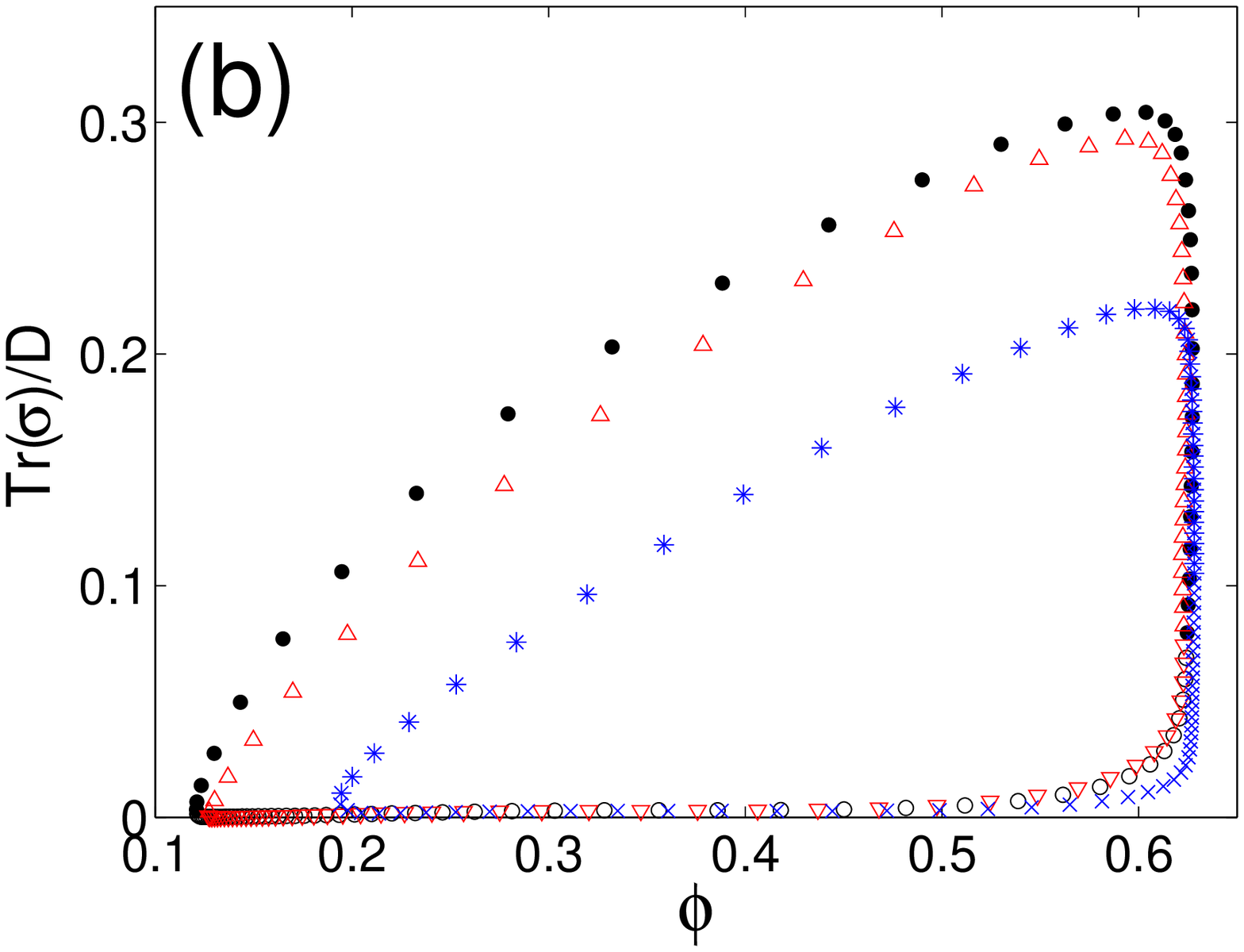}
\includegraphics[width=.6\columnwidth]{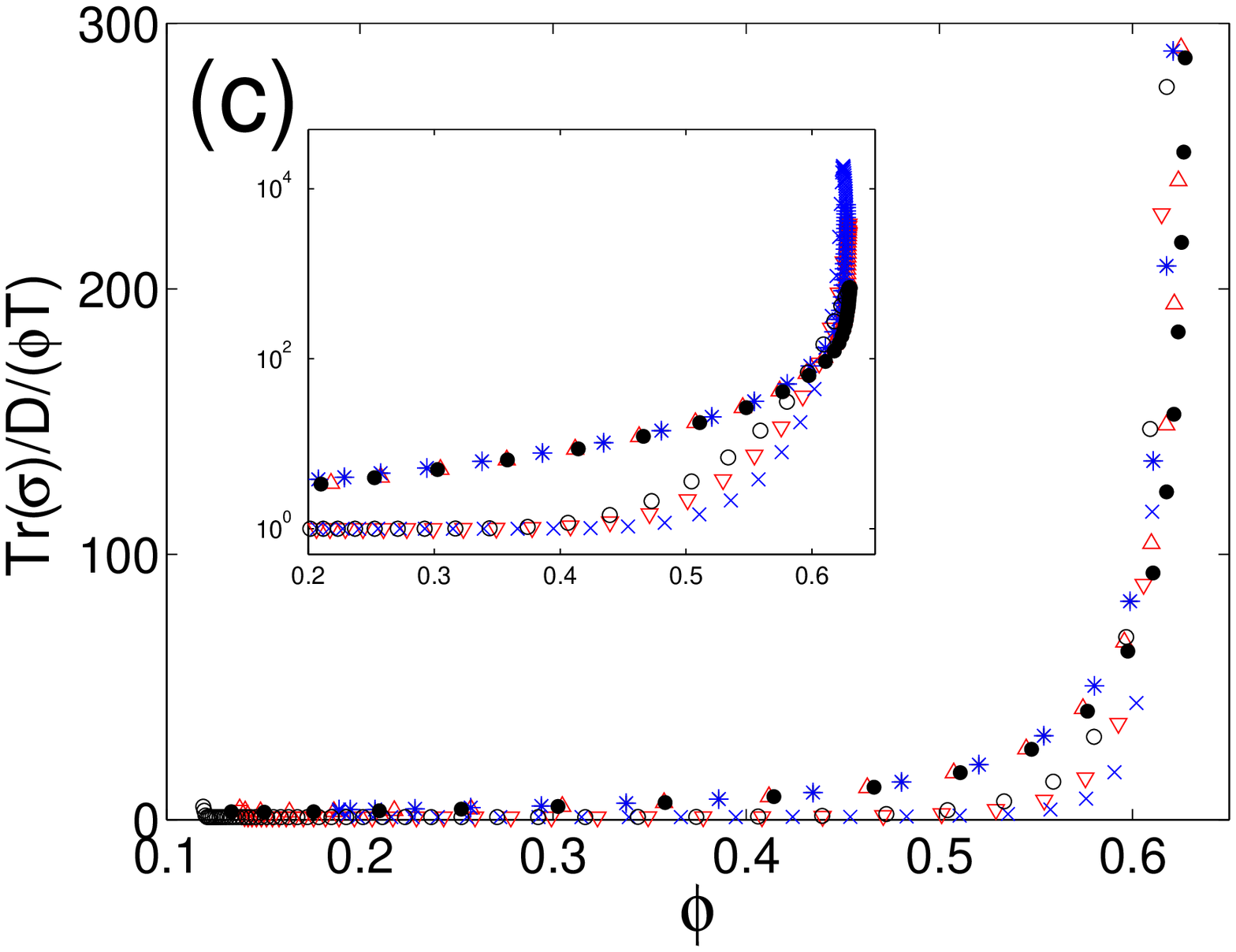}
\end{center}
\caption{\label{path_stressVFs}
(Color online)
(a) The granular temperature, (b) average normal stress, and (c)
average normal stress divided by the solid volume fraction and the
temperature
(i.e.  the unity plus inelastic dense gas correction in the equation
of state) for the three cases in Fig.~\ref{vfs}
($\bullet, \circ: \phi_{avg} = 0.36$;
$\bigtriangleup, \bigtriangledown: \phi_{avg} = 0.44$;
$\ast, \times: \phi_{avg} = 0.52$).
Measurements from the compaction region where particles accumulate
(characterized by $du_z/dz < 0$) are distinguished from those
from the dilation region where particles rain down ($du_z/dz > 0$);
$\bullet, \bigtriangleup, \ast$'s are obtained from the compaction
regions, and $\circ, \bigtriangledown, \times$'s are obtained
from the dilation regions.
The same symbols will be consistently used through
Fig.~\ref{path_bviscosityVFs},
and all quantities are dimensionless.
}
\end{figure*}

We start by plotting the scalar granular temperature $T$ and
average normal stress ${1\over D}{\rm Tr}(\uuline{\sigma})$
against the volume fraction for the three cases in Fig.~\ref{vfs}
[see Figs.~\ref{path_stressVFs} (a) and (b); in (b), the
volume fraction in the plateau region remains the same for
a range of solid phase stresses, and the exact distinction
between dilation and compaction branches in this region is elusive].
The computed values in the dilation region are nearly the
same for all three cases; however, those in the compaction
region strongly depend on $\phi_{avg}$.
Furthermore, they are significantly different from (and much
larger than) those in the dilation region.
It is clear that both $T$ and ${1\over D}{\rm Tr}(\uuline{\sigma})$
lie on two distinct branches, depending on where the measurements
of these variables are made.
Similar ``path dependence'' was also observed in lattice-Boltzmann
simulations of liquid-fluidized beds~\cite{derksen}.

In the previous section, we have shown that the granular
temperature tensor manifests significant anisotropy, which
has to be accounted for, in principle, in our analysis.
However, an appreciable anisotropy is seen only in small portion
of the compaction region [see inset of Fig.~\ref{vfs} (b)],
while the anisotropy [Fig.~\ref{vfs} (b)] and path dependence
[Fig.~\ref{path_stressVFs} (b)] in the stress tensor are
manifest over a wider region.
Thus, it is reasonable to believe that neither the anisotropy nor
the path dependence of the stress arise from the anisotropy of the
granular temperature.
Therefore, in what follows, we consider only the scalar granular
temperature.

For 1D flows:
%The solid phase stress for 1D flows reads:
%The relation for the solid phase stress for 1D flows:
\begin{equation}
 \label{mu_b}
{1\over D}{\rm Tr}(\uuline{\sigma}) = p_s -\mu_b {du_z\over dz},
\end{equation}
where $p_s$ is the dimensionless solid phase pressure,
and $\mu_b$ is the dimensionless bulk viscosity.
According to KTGM,
\begin{equation}
 \label{eos}
p_s = \phi T[1 + 2(1+e)\phi g_0(\phi)],
\end{equation}
where $g_0(\phi)$ is the isotropic ``equilibrium'' radial
distribution function (RDF) evaluated at contact.
We expect that a part of the path dependence of
${1\over D}{\rm Tr}(\uuline{\sigma})$ arises from the path
dependence of $T$.
We examine whether the first term in Eq.~(\ref{mu_b})
alone (i.e. neglecting the bulk viscosity effect) is enough to
explain the path dependence of the stress.
We rescale the average normal (dimensionless) stress by $\phi T$
in Fig.~\ref{path_stressVFs} (c), which corresponds to the 
inelastic and dense gas correction (plus unity) in KTGM, assuming
that the relation in Eq.~(\ref{eos}) holds.
We find that the path dependence still persists. Therefore,
Eq.~(\ref{eos}) cannot adequately explain the path dependence.

\begin{figure}[b]
\begin{center}
\includegraphics[width=.7\columnwidth]{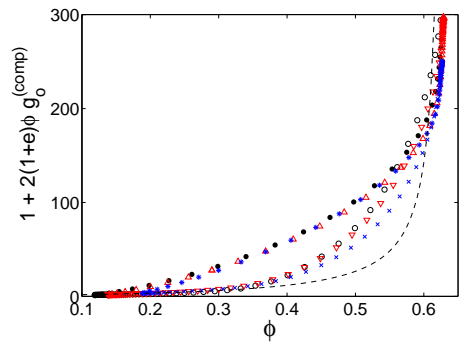}
\end{center}
\caption{\label{rdf}
(Color online)
Computationally obtained isotropic RDF at contact
(see text for details),
for the three cases in Fig.~\ref{path_stressVFs},
which are appropriately scaled for a better comparison with
Fig.~\ref{path_stressVFs} (c).
The RDFs indeed exhibit path dependence.
The dashed line is a scaled plot of an analytical formula used
by Bagnold, $g_o(\phi) = 1/[1-(\phi/\phi_{max})^{1/3}]$, where
$\phi_{max} = 0.63$ ($\sim$ the plateau volume fraction value)
was used.
}
\end{figure}

Physically, the path dependence of the average normal stress can be
rationalized as follows: As a granular assembly gets dilated or
compacted, the RDF at contact departs from its equilibrium value,
and depends on the rate of dilation/compaction.
When the relevant dimensionless group, ${du_z\over dz}/\sqrt{T}$,
is negligibly small, the assembly equilibrates quickly and remains
close to the equilibrium configuration.
(Note that, in this study, all the results are presented in terms
of dimensionless quantities. Since we use the particle diameter as
the characteristic length, this dimensionless group becomes
${du_z\over dz}/\sqrt{T}$ when $u_z$, $z$, and $T$ are all
dimensionless. In terms of dimensional variables, this group would
be $d_p{du_z\over dz}/\sqrt{T}$.)
However, when this quantity is somewhat larger, the RDF at contact
is expressed as a perturbation from its equilibrium value, and the
bulk viscosity term accounts for the departure from the equilibrium
value.
[In such a perturbation approach, $g_0$ in Eq.~(\ref{eos}) will
be the equilibrium value and is only a function of particle volume
fraction.]
Therefore, there is some rationale in trying to explain
the path dependence as the effect of the bulk viscosity.

We ascertain that the actual RDF at contact is indeed
path-dependent by directly computing it from our simulations.
The isotropic RDF at distance $r = r_n$ is~\cite{rapaport04}:
\begin{equation}
g(r_n) = {N_{shell}\over n_{avg} 4\pi r_n^2\Delta r},
\end{equation}
where $r_n = (n-1/2)\Delta r$, $n_{avg}$ is the average
number density, and $N_{shell}$ is the number of particles
lying in a thin spherical shell bounded by $(n-1)\Delta r$
and $n\Delta r$.
We compute the first five values of the RDF ($1 \leq n \leq 5$,
using $\Delta r = 10^{-2}$), and extrapolate them to estimate
its value at contact $r = 0.5$ (Fig.~\ref{rdf});
for the sake of better comparison with the results in 
Fig.~\ref{path_stressVFs} (c), a prefactor of $2(1+e)\phi$
is multiplied and unity is added.
The results vary a little bit depending on the choice of
$\Delta r$ and the extrapolation method.
However, in all cases, the path dependence persists.
Thus we can indeed expect a non-negligible bulk viscosity correction.
Before discussing the bulk viscosity effect, let us look at the shear
viscosity.

\begin{figure}[t]
\begin{center}
\includegraphics[width=.7\columnwidth]{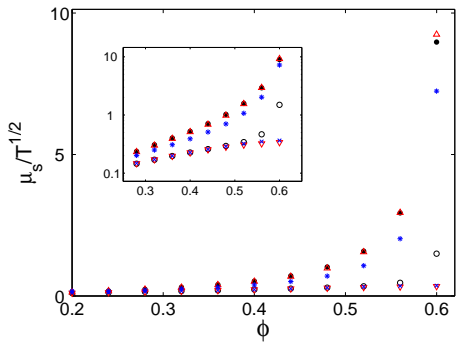}
\end{center}
\caption{\label{path_sviscosityVFs}
(Color online)
Solid phase dimensionless shear viscosity scaled by dimensionless
$\sqrt{T}$, obtained for the three different values of $\phi_{avg}$
in Fig.~\ref{path_stressVFs}.
Insets include the viscosities shown in linear-logarithmic scale.
}
\end{figure}

We numerically estimate the first derivative of the average velocity
fields through the central differencing scheme, and compute the
effective (dimensionless) shear viscosity $\mu_s$ from different
components of the stress tensor:
\begin{equation}
 \label{mu_s}
\sigma_{zz} - {1\over D}{\rm Tr}(\uuline{\sigma}) = -{4 \over 3}\mu_s{du_z\over dz}.
\end{equation}
As the shear viscosity in KTGM depends on $\sqrt{T}$, we plot
$\mu_s$ after rescaling it by $\sqrt{T}$ (Fig.~\ref{path_sviscosityVFs}).
We find that scaled $\mu_s$'s computed from different $\phi_{avg}$'s
collapse nearly onto common functional forms, but the path dependence
is very clear.
In the KTGM, the shear viscosity has no path dependence~\cite{gidaspow94}:
\begin{eqnarray}
 \label{mu_sKTGM}
\mu_s & = & {5\sqrt{\pi}\over 48(1+e)g_0}\sqrt{T}\left[1+{4\over 5}(1+e)\phi g_0\right]^2 \nonumber \\
&& + {4\over 5}\phi^2g_0(1+e)\sqrt{{T\over \pi}},
\end{eqnarray}
which holds, again, for small magnitude of ${du_z\over dz}/\sqrt{T}$,
and the apparent path dependence again signals clearly a marked
departure from conditions assumed in the development of KTGM.

\begin{figure}[t]
\begin{center}
\includegraphics[width=.63\columnwidth]{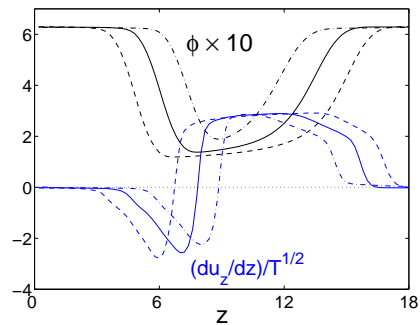}
\end{center}
\caption{\label{dudz}
(Color online)
A dimensionless group relevant to the rate of dilation or
compaction, ${du_z\over dz}/\sqrt{T}$, for the cases in
Fig.~\ref{vfs} (dashed lines: $\phi_{avg} = 0.36$;
solid lines: $\phi_{avg} = 0.44$;
dot-dashed lines: $\phi_{avg} = 0.52$),
plotted together with the scaled volume fraction profile.
[Here $z,~u_z$, and $T$ are all dimensionless.]
}
\end{figure}

We have computed the dimensionless quantity ${du_z\over dz}/\sqrt{T}$
from our simulation results; see Fig.~\ref{dudz}.
Note that the magnitude of this quantity is not small compared to
unity, and so the assumption that the actual RDF at contact can be
written as a small perturbation from the equilibrium value, which is 
routinely made in the theories, does not hold in this rather simple
inhomogeneous flow.
The path dependence of the shear viscosity is thus clearly associated
with effects which are nonlinear in ${du_z\over dz}/\sqrt{T}$. 

\begin{figure}[t]
\begin{center}
\includegraphics[width=.7\columnwidth]{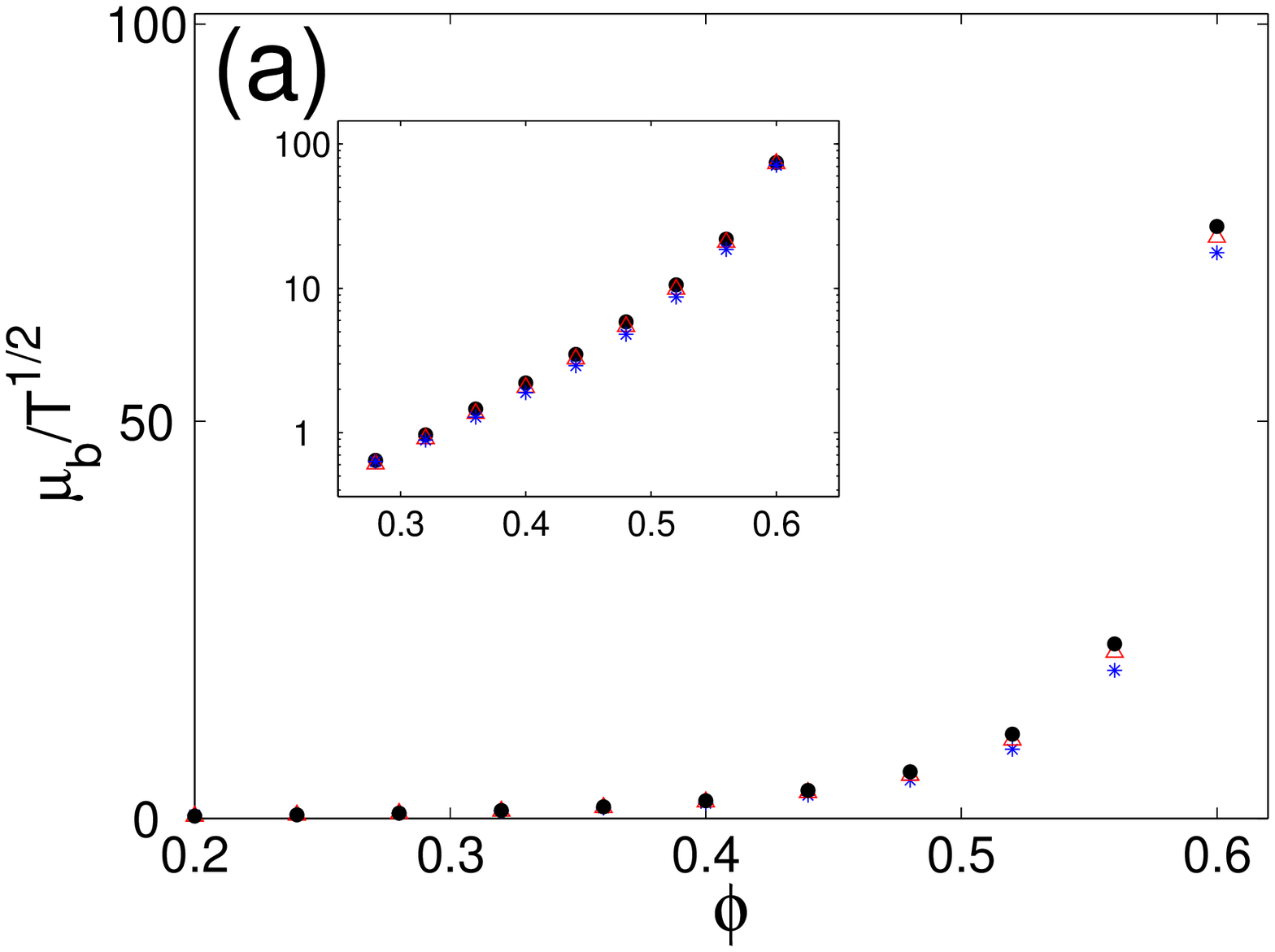}
\includegraphics[width=.7\columnwidth]{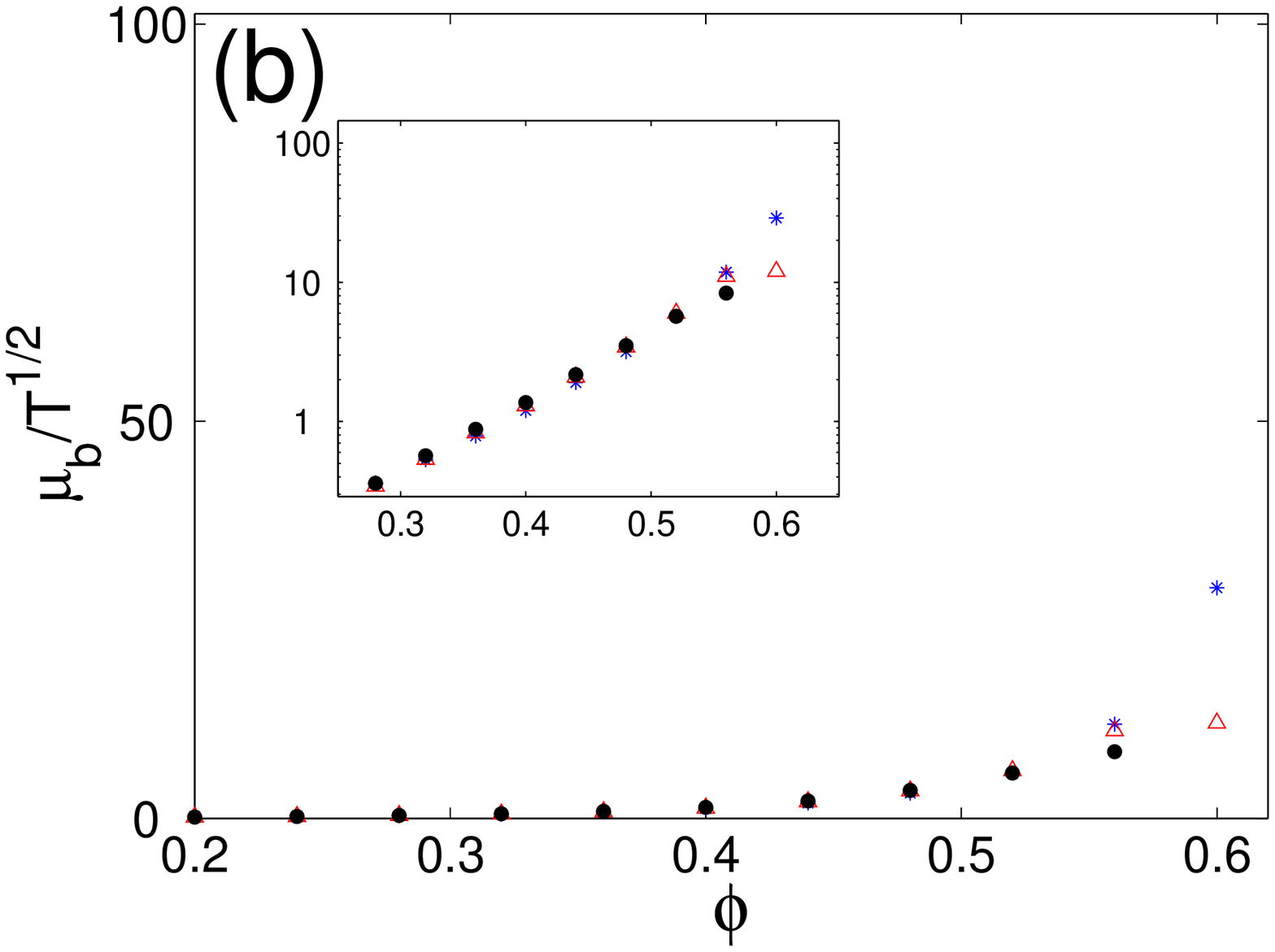}
\end{center}
\caption{\label{path_bviscosityVFs}
(Color online)
Solid phase dimensionless bulk viscosity scaled by dimensionless
$\sqrt{T}$, obtained for the three different values of $\phi_{avg}$
in Fig.~\ref{path_stressVFs},
estimated via (a) Eq.~(\ref{compute}), and (b) Eq.~(\ref{compute2}).
Insets include the viscosities shown in linear-logarithmic scale.
}
\end{figure}

Having found that $\mu_s$ is path-dependent, we can expect the bulk
viscosity $\mu_b$ to differ in the compaction and dilation branches
as well.
Strictly speaking, there is no unique way of partitioning
${1\over D}{\rm Tr}(\uuline{\sigma})$ into $p_s$ and
$\mu_b{du_z\over dz}$, if $\mu_b$ is going to depend on $\phi$,
$T$, and the path.
However, it is clear from Fig.~\ref{path_stressVFs} that $T$ and
${1\over D}{\rm Tr}(\uuline{\sigma})$ in the dilation branch
are nearly identical for the three cases, and independent of
${du_z\over dz}/\sqrt{T}$.
This suggests that the bulk viscosity in the dilation branch is
close to zero, so that ${1\over D}{\rm Tr}(\uuline{\sigma})$
in the dilation branch can be taken as simply equal to $p_s$
in that branch.
As seen in Fig.~\ref{path_sviscosityVFs}, the shear viscosity
in the dilation branch is much smaller than that in the compaction
branch. Thus, it is not unreasonable to suspect that the bulk
viscosity correction in the dilation branch is small:
\begin{equation}
p_s|_{dil} \approx {1\over D}{\rm Tr}(\uuline{\sigma})|_{dil}, \nonumber
\end{equation}
where the subscript $dil$ indicates evaluation on the dilation
branch.
Armed with this, we can estimate the bulk viscosity in the
compaction branch in at least two different ways. First, we
can set (assume) $p_s|_{dil} \approx p_s|_{comp}$ and write
\begin{equation}
 \label{compute}
{1\over D}{\rm Tr}(\uuline{\sigma})|_{comp} = {1\over D}{\rm Tr}(\uuline{\sigma})|_{dil} - \mu_b{du_z\over dz}|_{comp},
\end{equation}
where the subscript $comp$ indicates evaluation on the compaction
branch.
Bulk viscosity values in the compaction branch estimated in this
manner are shown in Fig.~\ref{path_bviscosityVFs} (a).
Alternately one can recognize that the granular temperature is
different in the two branches; accounting for the difference,
we can write
\begin{equation}
 \label{compute2}
{1\over D}{\rm Tr}(\uuline{\sigma})|_{comp} = {1\over D}{\rm Tr}(\uuline{\sigma})|_{dil}\times\left({T_{comp}\over T_{dil}}\right) - \mu_b{du_z\over dz}|_{comp}.
\end{equation}
Figure \ref{path_bviscosityVFs} (b) shows the bulk viscosity values
estimated via Eq.~(\ref{compute2}).
In both Figs.~\ref{path_bviscosityVFs} (a) and (b), the bulk viscosity
values are scaled by $\sqrt{T_{comp}}$. The insets in these figures
show the results in a linear-logarithmic scale.
The second approach leads to smaller estimates for the bulk viscosity.
However, it is clear that {\it irrespective} of the approach (assumption)
taken [i.e. Eq. (\ref{compute}) or (\ref{compute2})], the estimated values
of bulk viscosity in the compaction branch are found to be appreciably
larger than the shear viscosity (compare Figs. \ref{path_sviscosityVFs}
and \ref{path_bviscosityVFs}).
It is interesting to contrast this with KTGM
predictions~\cite{gidaspow94}:
\begin{equation}
\mu_b = {4\over 3}\phi^2g_0(1+e)\sqrt{{T\over\pi}}
\end{equation}
which is smaller than the magnitude of the shear viscosity.

It should be emphasized that the 1D-TW studied here is not a
peculiar problem representing extreme conditions.
Such sharp volume fraction gradients routinely occur around
bubble-like voids and clusters in fluidized beds, turbulent
beds, and fast fluidized beds~\cite{davidson85}.
In such devices, the particle assembly is frequently subjected
to local dilation and compaction in alternating manner.
In these flows, $d_p\nabla\cdot {\mathbf u}/\sqrt{T}$
may not be small and this can lead to the type of effects seen
in this study.

\subsection{Effect of cohesion}
 \label{cohesion}

It is well known that beds of fine powders are difficult to
fluidize because of the cohesive force between the particles.
Beds of cohesive powders have attracted increased attention
in recent years; however, their theoretical understanding
is still limited.
In this section, we vary the level of cohesion to probe
how the cohesive force changes the wave profiles and the
relations among continuum variables.

\begin{figure*}[t]
\begin{center}
\includegraphics[width=.7\columnwidth]{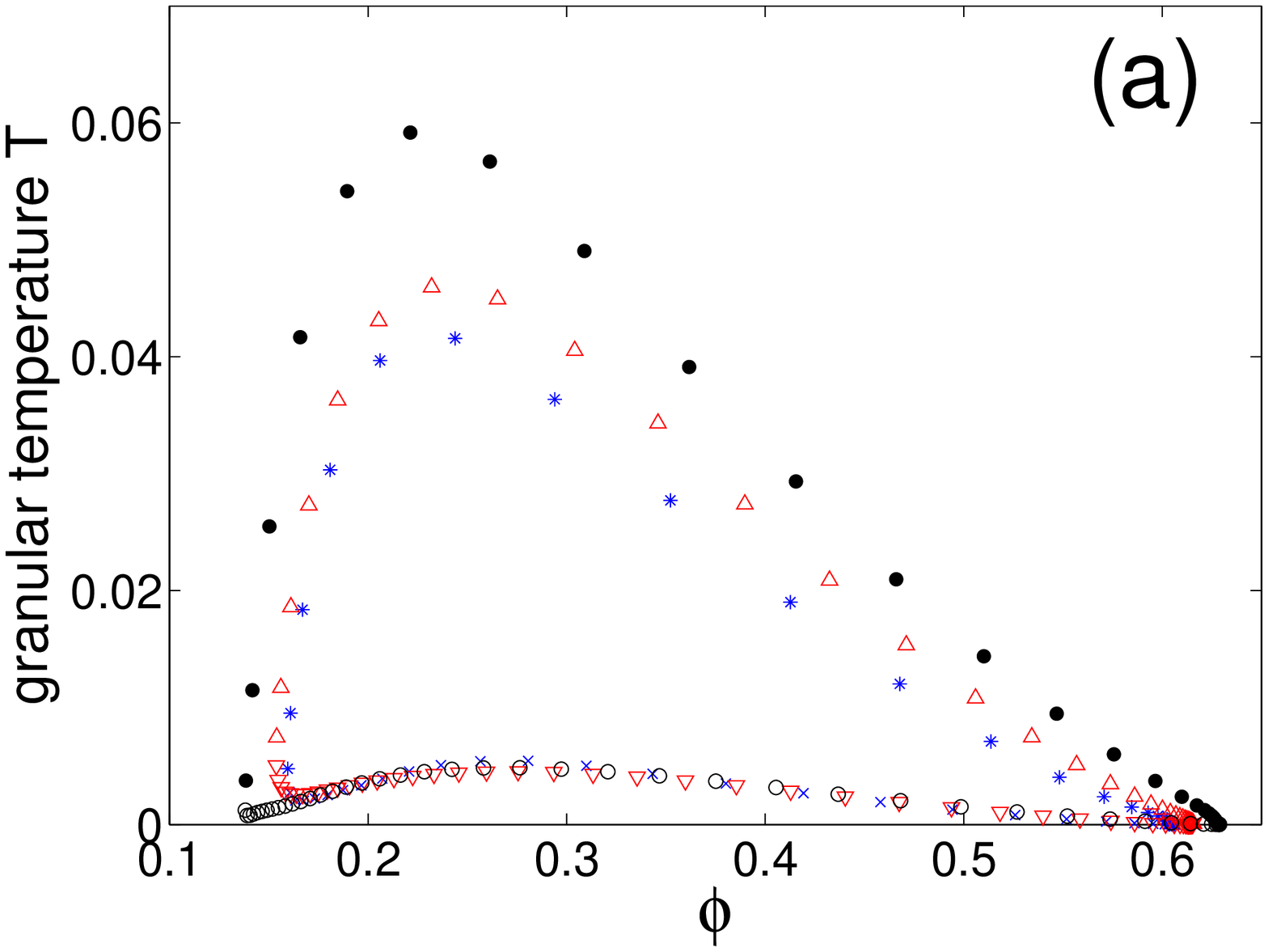}
\includegraphics[width=.7\columnwidth]{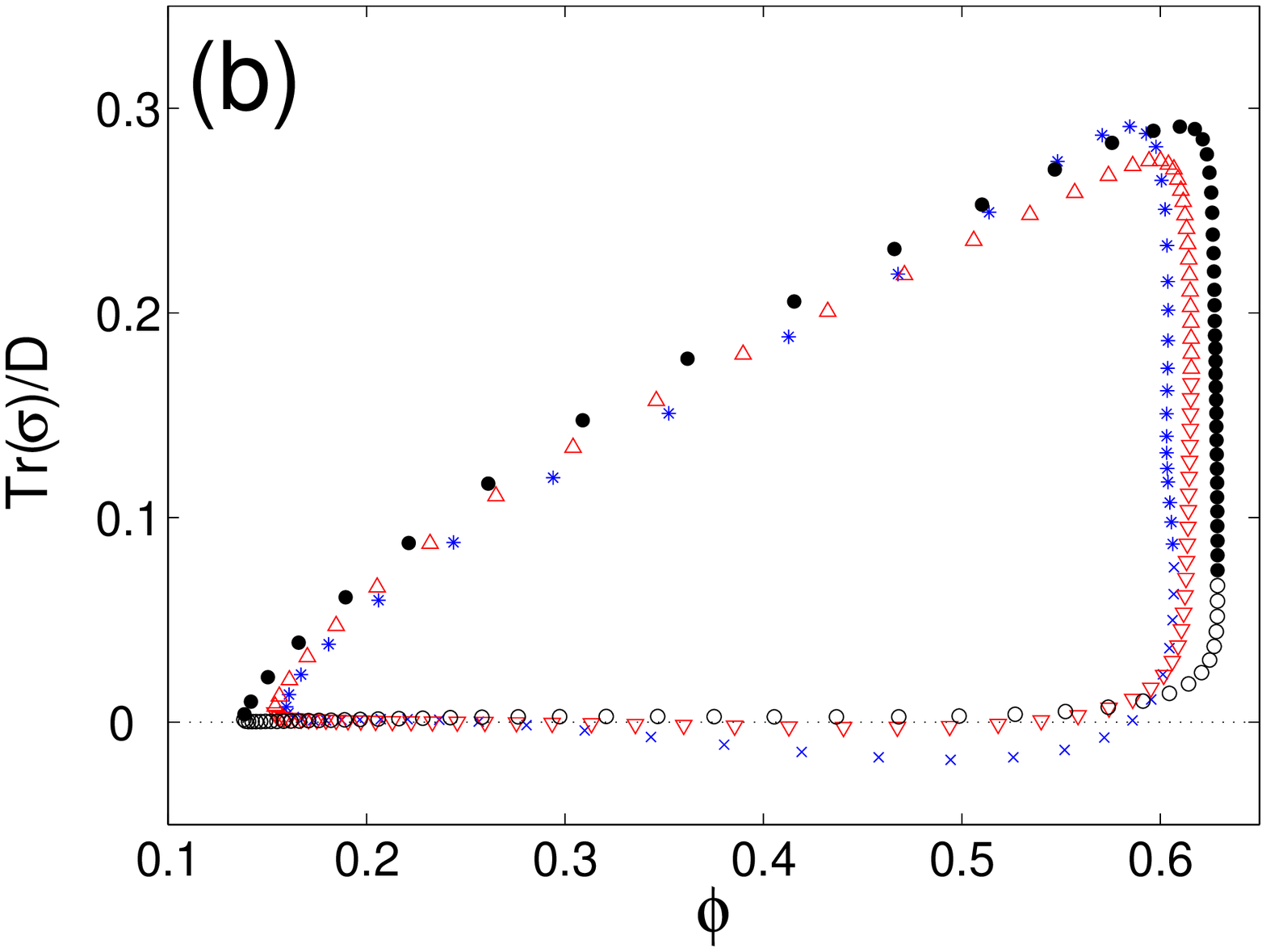}
\end{center}
\caption{\label{path_stressBos}
(Color online)
Solid phase granular temperature and average normal stress for
three cases of different cohesion level ($\bullet, \circ: Bo = 0$;
$\bigtriangleup, \bigtriangledown: Bo = 1$;
$\ast, \times: Bo = 2$); $\phi_{avg} = 0.44$ in all cases
($e = 0.9, \mu = 0.1$).
$\bullet, \bigtriangleup, \ast$'s are obtained from the compaction
regions, and $\circ, \bigtriangledown, \times$'s are obtained
from the dilation regions.
($St = 55$.)
The wave speed monotonically slows down with increasing level of
cohesion, until the wave disappears at $Bo \sim 8$;
the wave speed in beds of $Bo = 2$ is $\sim 90\%$ of that in
$Bo = 0$.
}
\end{figure*}

\begin{figure}[b]
\begin{center}
\includegraphics[width=.7\columnwidth]{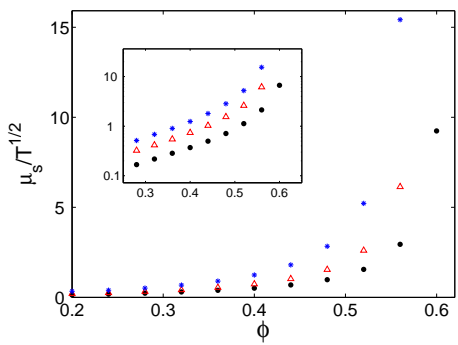}
\end{center}
\caption{\label{path_sviscosityBos}
(Color online)
Solid phase dimensionless shear viscosity in the compaction branch
($\phi_{avg} = 0.44$), scaled by dimensionless $\sqrt{T}$,
obtained from the three cases of different values of $Bo$ in
Fig.~\ref{path_stressBos}
($\bullet: Bo = 0$; $\bigtriangleup: Bo = 1$; $\ast: Bo = 2$.)
The viscosities are shown in linear-logarithmic scale in the inset.
}
\end{figure}

The waves formed in beds of weakly cohesive particles ($Bo <\sim 2$)
retain a virtually invariant shape over time, and travel with a
well-defined constant speed.
They become more and more irregular and slow down, until they vanish
at $Bo \sim 8$, depending on $\lambda$ and $\phi_{avg}$;
the propagation speed at $Bo = 6~(3)$ is about 60\% (85\%)
of that in beds of $Bo = 0$~\cite{moon05c}.
The wave profiles for non-cohesive and weakly cohesive particles
($Bo \leq 2$) are shown in (Fig.~\ref{bos}).
There are two major differences between the waves in beds of
non-cohesive and of cohesive particles:
Firstly, the plateau volume fraction is smaller in cohesive beds,
as cohesive particles pack more loosely than non-cohesive ones,
due to reasons similar to those for frictional particles
(See e.g., Dong et al.~\cite{dong06}).
Even for slightly cohesive particles, this effect is appreciable
(see inset of Fig.~\ref{bos}); i.e. the effect is more pronounced
compared to that of a changing friction coefficient (Fig.~\ref{ideal}).
Secondly, the average normal stress becomes tensile and remains
so for a certain range of the height in the dilation region.
At the bottom of the upper plug, the average normal stress
monotonically decreases until it reaches its minimum value
(the largest tensile stress) and the assembly breaks up
(inset in Fig.~\ref{bos}).
The magnitude of the tensile stress cannot exceed the tensile
strength of the cohesive assembly; thus, one can expect that
the former corresponds to the tensile strength of this
fluidized bed.
When we estimate the tensile strength via this reasoning,
it is almost an order of magnitude smaller that what is predicted
by Rumpf's model~\cite{moon05c}.
%It is understandable,
This makes sense, since the cohesive assembly in a fluidized
bed breaks up through the direction of the weakest linkage,
in contrast to all directions isotropically, as assumed in
Rumpf's model~\cite{rumpf62}.
We observe further decrease (increase) of the plateau volume fraction
(the magnitude of maximum tensile strength), as $Bo$ increases,
until the waves vanish (i.e. the bed does not get fluidized any more)
at $\sim 8$.
The size of agglomerates broken off of the cohesive assembly
however has a distribution, and the wave profile becomes irregular
for $Bo >\sim 3$, which makes an quantitative analysis difficult.
For sufficiently cohesive particles ($Bo > \sim 8$),
as one can readily expect, the maximum tensile stress driven
solely by gravity fails to reach the tensile strength of the
material, and the bed does not form a stable traveling wave
any more.
At that point, an additional mechanism, such as mechanical
vibration or fine powder coating, has to be implemented to
fluidize them~\cite{nam04,mohan04}.

\begin{figure*}[t]
\begin{center}
\includegraphics[width=.7\columnwidth]{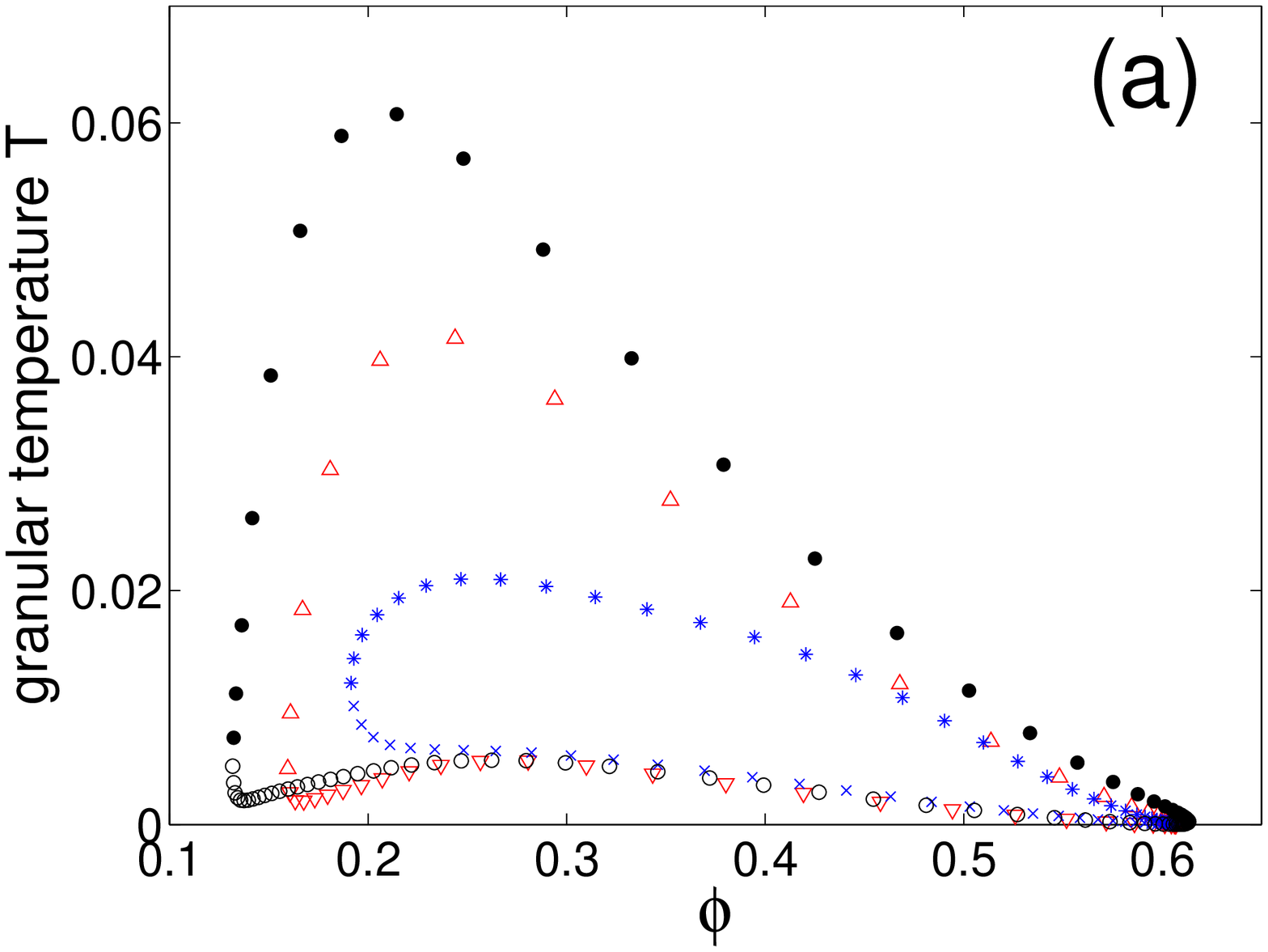}
\includegraphics[width=.7\columnwidth]{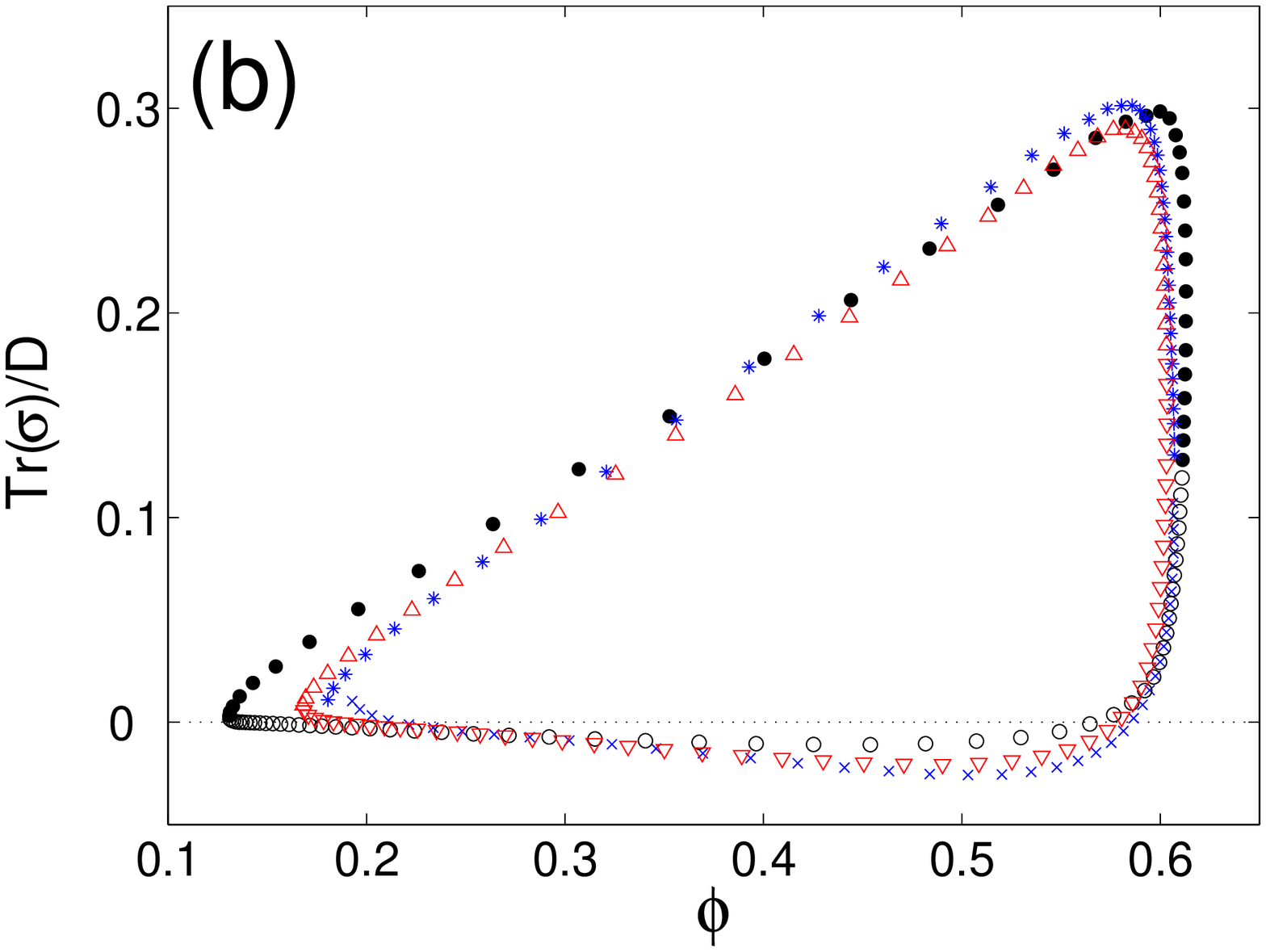}
\end{center}
\caption{\label{path_stressBoVFs}
(Color online)
Solid phase granular temperature and average normal stress in beds
of cohesive particles
($Bo = 2; e = 0.9; \mu = 0.1$) for three different average volume
fractions ($\bullet, \circ: \phi_{avg} = 0.40$;
$\bigtriangleup, \bigtriangledown: \phi_{avg} = 0.44$;
$\ast, \times: \phi_{avg} = 0.48$).
$\bullet, \bigtriangleup, \ast$'s are obtained from the compaction
regions, and $\circ, \bigtriangledown, \times$'s are obtained
from the dilation regions.
($St = 55$.)
}
\end{figure*}

As in the beds of non-cohesive particles, both $T$ and
Tr$(\uuline{\sigma})/D$ form lobes when plotted against $\phi$,
and manifest the same path dependence
(Figs.~\ref{path_stressBos} and \ref{path_stressBoVFs}).
Clearly, our arguments on pronounced bulk viscosity effects,
and on path dependence of both the bulk and shear viscosities
inferred for beds of non-cohesive particles should remain valid
for beds of moderately cohesive particles as well.
As $Bo$ increases (in a narrow range of $Bo \leq 2$, at a
fixed average volume fraction), both the maximum granular
temperature [Fig.~\ref{path_stressBos} (a)]
and the maximum volume fraction in the plateau region
[Fig.~\ref{path_stressBos} (b)] decrease noticeably, but
the maximum (compressive) value attained by the average normal
stress remains nearly the same.
We compute the shear viscosity for the three cases shown in
Fig.~\ref{path_stressBos}, and find that it increases as $Bo$
increases (only the values on compaction branches are shown in 
Fig.~\ref{path_sviscosityBos} for clarity), as one can expect.

For a fixed level of cohesion ($Bo = 2$), the maximum granular
temperature decreases as the average volume fraction increases
[Fig.~\ref{path_stressBoVFs} (a)],
because the void region gets narrower and particles accumulate
more gently as in the beds of non-cohesive particles with
varying average volume fraction.
The average normal stress value for different $\phi_{avg}$
in the compaction branch varies.
However, the granular temperature [Fig.~\ref{path_stressBoVFs} (a)]
and average normal stress [Fig.~\ref{path_stressBoVFs} (b)]
in the dilation branch for the three cases follow (roughly) the same
curve until the assembly breaks off;
as in beds of non-cohesive particles, dilation branches follow
nearly the same curves.

\section{Discussion}
 \label{conclusion}

Soft sphere models for particle-particle interactions have been
used in the literature for nearly three decades to analyze a
variety of dense particulate flow problems~\cite{cundall79,hopkins90,
mehta91,mccarthy96,potapov96,cleary01,campbell02,cleary02,landry03,zhu04}.
The particle dynamics-based hybrid model employed in our study,
which combines a soft-sphere model for the particles with
local-average equations for the gas phase, has been used
previously to simulate flows of gas-particle mixtures~\cite{tsuji93,
hoomans96,xu97,kawaguchi00,mcnamara00,rhodes01a,zhou04,ye04,feng04},
including those occurring in multi-dimensional fluidized
beds~\cite{tsuji93,hoomans96,xu97,rhodes01a,zhou04,ye04,feng04},
and successfully compared against experimental
data~\cite{vanWachem01,hoomans01}.
Thus there is ample basis for using this approach to probe
the nature of inhomogeneous gas-particle flows.
We have used this model to simulate one of the simplest
spatiotemporal structures possible in gas-fluidized beds,
namely one-dimensional traveling waves (1D-TW).
We believe that such a simple structure is ideal for our study,
as it contains both compacting and dilating regions, and
its steady nature (in the co-traveling frame) allows us to
perform suitable ensemble-averaging, which is essential for
the analysis of results of discrete simulations.
Through an analysis of ``computational data'' generated through
simulations of these traveling wave structures, we have examined
in this study the possible consequences of compaction and dilation
rates on the stresses generated in the particle assembly.
Such volume changes accompany most gas-particle flows, and
understanding their effects on the stresses is thus of practical
importance.

It is well established in the literature, through stability analysis
based on continuum two-fluid models, that such traveling waves are
the most dominant modes through which homogeneously fluidized
suspensions lose stability.
In real gas-fluidized beds, such traveling waves are not observed,
as they quickly give way to a secondary instability leading to
bubble-like voids~\cite{anderson95,glasser96,sundar03}.
However, one can readily generate such traveling waves in computations
by restricting the cross-section of the bed to only several particle
diameters, thereby suppressing the secondary instability~\cite{duru02}.
Although the 1D-TW generated in this manner are artificial structures,
they represent a simple, suitable inhomogeneous flow pattern where the
particle assembly undergoes compaction and dilation alternately in
a periodic manner.

We began by generating computational data on 1D-TW for assemblies
of uniformly sized, spherical, non-cohesive particles.
We demonstrated that the particle volume fraction profiles observed
in the hybrid model calculations are qualitatively similar to those
seen previously in analyses of two-fluid models~\cite{anderson95,glasser96}.
These traveling waves in dense fluidized beds take the form of
vertically rising plugs of particles.
The particle volume fraction in the plugs was found to be essentially
independent of the average particle volume fraction in the wave,
while the height of the void and the lowest volume fraction attained
in this void depended systematically on it
(see Figs.~\ref{vfs} and \ref{lambda}) --- just as
in the case of similar structures obtained via two-fluid models
supplemented with {\it ad hoc} closures~\cite{anderson95,glasser96}.
The present hybrid model replaces the {\it ad hoc} closures for
solid phase stresses in the two-fluid model with particle-particle
interaction models.

The variation of wave amplitude with wavenumber (Fig.~\ref{amps}) obtained
in our simulations is also similar to that obtained with two-fluid models.
Typically, in two-fluid models, the homogeneous state is stable for
short wavelengths, and it gives way to traveling waves via a Hopf
bifurcation at some critical wavelength.
As the wavelength increases further, the wave amplitude
increases~\cite{glasser96}.
Although the discrete simulations undertaken in the present study
do not exhibit a perfect bifurcation from a uniform solution to a
traveling wave solution, one can readily recognize the correspondence
between our discrete simulations and corresponding two-fluid model
calculations of Glasser et al.~\cite{glasser96}.

Our discrete simulations clearly show that the solid phase stresses
and the granular temperature obtained by ensemble-averaging over many
realizations are path-dependent.
We have checked that the normal stress difference in bubbling fluidized
beds is different from the well-known Burnett order effect in sheared
granular fluids (See Appendix).
The traveling wave can be partitioned into two regions, one where the
assembly undergoes dilation and the other where it undergoes compaction.
The solid phase stresses and the granular temperature in these
two regions differ markedly (see Fig.~\ref{path_stressVFs}).
While one does expect some difference in the behavior observed in
the two regions, our simulations reveal that the difference is much
larger than what can be attributed to the bulk viscosity correction
in the widely used kinetic theory for granular materials (KTGM).
The kinetic theory is limited to situations where the volumetric strain
rate is small compared to the rate at which local microstructure
equilibrates, so that the effect of the volumetric strain rate can be
treated as a small perturbation from the equilibrium microstructure.
Our computations show that, in this simple traveling wave, the
volumetric strain rate is not small (see Fig.~\ref{dudz}).
Although these traveling waves are not observed in real gas-fluidized
beds, it is such one-dimensional traveling waves that give way
to bubble-like voids, and so one can readily expect that the same
effects will be observed in bubble-like voids as well. 

Large (scaled) volumetric strain rates can be expected to lead to
nonlinear effects, and this is indeed seen.
Both bulk and shear viscosities in the solid phase are found
to depend on the volumetric strain rate (see Figs.~\ref{path_sviscosityVFs}
and \ref{path_bviscosityVFs}), and such dependence is not captured
by the KTGM (which only considers the case of small-scale volumetric
strain rates). 
When two-fluid models fail to account for such nonlinear effects,
they cannot be expected to reproduce the macroscopic properties of
the void region accurately.
It is well established that bubble-like voids readily form in
two-fluid model calculations even with a simple phenomenological
closure for the particle phase stress~\cite{gidaspow94,anderson95,glasser96};
however, the shape of such voids is not consistent with those
observed experimentally.
A part of this problem stems from numerical errors associated with
discretization of the two-fluid model equations~\cite{guenther01};
however, numerical accuracy is not the sole factor.
In the dense emulsion phase in the immediate vicinity of the
bubble-like void, one can expect large-scale volumetric strain rates
(just as in the 1D-TW examined in this study)
and the failure to account for the accompanying nonlinear effects
will impact the characteristics of the bubble-like voids.

In many technological applications involving gas-particle flows,
particles in the 50 - 100 $\mu$m size range are employed.
Such particles, belonging to Geldart type A~\cite{geldart73},
manifest modest levels of inter-particle cohesion and the relevance
of this cohesive interaction on the flow characteristics has been
debated extensively in the literature (for example, see a review
article by Sundaresan~\cite{sundar03});
however, concrete conclusions are yet to emerge.
In the present study, we have also examined the effect of inter-particle
cohesion on the structure of the traveling wave.
The inter-particle cohesive force model itself is quite simple,
but it suffices to explore the generic effects.
It is found that all of the effects observed in non-cohesive
systems persist when a modest level of cohesion is added
(see Figs.~\ref{path_stressBos} and \ref{path_stressBoVFs}).
Cohesion lowers the particle volume fraction in the plateau region
and hence the amplitude of the wave.
It also slows down the speed of the wave propagation (see caption
of Fig.~\ref{path_stressBos}) --– which (in multi-dimensional
systems) affords greater exchange of gas between the void and the
surrounding emulsion.
Thus the beneficial effect of modest levels of cohesion can be
understood. 

Although we have presented results only for $Bo$ up to 2, we found
that wave solutions could be obtained even for larger $Bo$ values.
At larger $Bo$ values (e.g., $Bo$ = 4), the waves became irregular
and when $Bo > \sim$ 8, the entire assembly traveled as a single plug.
This is not a surprising result, as it is well known that highly
cohesive particles (Geldart type C~\cite{geldart73}) do not fluidize well.
Thus the range of $Bo$ over which beds can be fluidized without
additional excitation (such as mechanical vibrations) is $Bo < \sim 8$.
Since the wave propagation is irregular for $Bo > \sim 3$,
the narrow range of $0 \leq Bo < \sim 3$ is more appropriate
for smooth fluidization and hence quantitative analysis
through ensemble-averaging.
At such levels of cohesion, the particle assembly is clearly in a state
of tension when it undergoes dilation (see Figs.~\ref{path_stressBos}
and \ref{path_stressBoVFs}). It also clearly speaks for the need to
include in two-fluid models a path-dependent model for the granular
pressure.

\section*{Acknowledgments}
S.J.M. is grateful to Lee Aarons for helpful discussions.
This work was supported by grants from The New Jersey Commission
on Science and Technology and Merck \& Co., Inc., the US D.O.E.,
and a Guggenheim Fellowship (I.G.K.).

\section*{Appendix: The normal stress differences in sheared
gas-fluidized beds}

Normal stress difference can arise due to Burnett order
effects (Sela and Goldhirsh~\cite{sela98}).
Theoretical prediction for normal stress difference
attributable to Burnett order effects is available
only for the {\it dilute} limit, and a validated Burnett
order hydrodynamical theory for bubbling fluidized beds
(i.e. {\it dense} suspensions) is currently unavailable.
We first estimated the normal stress differences
($\sigma_{zz}-\sigma_{xx}$) using a general relationship
provided by Sela and Goldhirsh~\cite{sela98}
(even though it was developed for the {\it dilute} limit),
and then considered a couple of relevant computational
experiments in order to probe if the stress differences that
we observed in our fluidized beds were consistent with Burnett
order effects.

\begin{figure}[t]
\begin{center}
\includegraphics[width=.7\columnwidth]{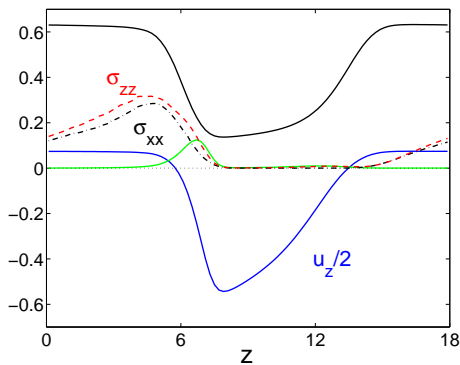}
\end{center}
\caption{
\label{shear_ideal}
Solid phase continuum variables of beds of ideal particles
($e = 1.0; \mu = 0.0; Bo = 0; St = 55; \phi_{avg} = 0.44$).
}
\end{figure}

According to a general Burnett order expression for the
stresses (Eq.~(46) in Sela and Goldhirsh~\cite{sela98}),
the normal stress difference in 1D-TW in fluidized beds
depends on (products of) gradients of solid phase continuum
variables, such as $du_z/dz$, $dT/dz$, and $d^2T/dz^2$.
We determined that, for the waves in our simulations, the
dominant contribution came from a term of the form
$C\phi\lambda^2(du_z/dz)^2$, where $C$ is a constant of order
of unity and $\lambda$ is the mean free path.
In {\it dense} suspensions, $\lambda$ is much smaller than unity.
We found this term to be much smaller than the
($\sigma_{zz}-\sigma_{xx}$) values that we obtained in our
simulations (see, for instance, Fig.~\ref{vfs} (b)).
Thus, there is no basis for attributing the stress difference
seen in our simulations to Burnett order effects.

To test this issue further, we also considered two additional
computational experiments:
%In order to probe the consistency between the normal stress
%differences in fluidized beds and Burnett order effects in
%sheared granular flows, we consider the following two
%computational experiments:
\begin{itemize}
\item[(i)]The Burnett order theory by Sela and
Goldhirsh~\cite{sela98} for homogeneously sheared granular flows
in the dilute limit showed that the normal stress differences
still persist in the elastic limit.
This theory also predicts that the {\it first} normal stress difference
($(\sigma_{zz}-\sigma_{xx})/({\rm Tr}(\uuline{\sigma})/D)$, in our case)
is a rapidly increasing function of inelasticity.
We first computed different diagonal components of the stress in
usual fluidized beds, using {\it ideal} particles
($e = 1.0, \mu = 0.0$) used in Fig.~\ref{ideal}, and examined
their dependence on the inelasticity.
\item[(ii)] Using dissipative particles ($e = 0.9, \mu = 0.1$),
we applied homogeneous shear in the $y-$direction to a fluidized bed;
if there were any systematic normal stress difference between the
$xx-$ and $yy-$ components, it would be mostly attributed to the
Burnett order effects (and/or collisional anisotropy~\cite{alam03}).
%; Alam and
%Luding~\cite{alam03} have recently shown that the first normal
%stress difference in a dense granular sheared flow arises from
%a different mechanism, i.e., collisional anisotropy).
\end{itemize}

In a bed of ideal particles (case (i) above, shown in
Fig.~\ref{shear_ideal}), a noticeable normal stress difference
($\sigma_{zz}-\sigma_{xx}$) persists at a level comparable
to the dissipative particle system.
Comparing the results for the ideal and dissipative systems,
we saw that the ratio
$(\sigma_{zz}-\sigma_{xx})/({\rm Tr}(\uuline{\sigma})/D)$
was not a strong function of inelasticity.
This further confirms that Burnett order effect is unimportant.
%(and {\it both relatively
%and absolutely} comparable to that of dissipative particles,
%as the average normal stress remains about the same)
%normal stress difference between $xx-$ and $zz-$ components
%still persists ($xx-$ and $yy-$ components are essentially
%the same within fluctuations because of symmetry).
%It clearly shows that the normal stress differences in
%bubbling fluidized beds have a different characteristic (that
%it is not a strong function of inelasticity).
%Note that we have another dissipation mechanism due to
%an interstitial gas.

\begin{figure}[t]
\begin{center}
\includegraphics[width=.7\columnwidth]{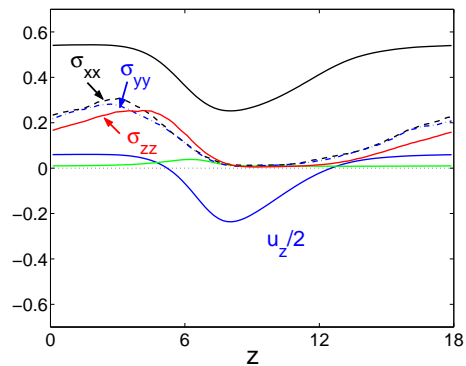}
\end{center}
\caption{
\label{shear_bed}
Solid phase continuum variables in sheared fluidized beds (computed
by employing Lees-Edwards boundary condition in the $y-$ direction).
The shear strain rate $du_x/dy$ of 0.2 is applied.
($e = 0.9; \mu = 0.1; Bo = 0; St = 55; \phi_{avg} = 0.44$).
}
\end{figure}

When we applied shear to a fluidized bed (case (ii), shown in
Fig.~\ref{shear_bed}) in the $y-$direction, we still saw
significant differences between $xx-$ (or $yy-$) and $zz-$ component
normal stresses, but those between $xx-$ and $yy-$ components
were relatively very small.
% (even though they seem to be systematic).
The maximum volumetric strain rate ($du_z/dz \sim 0.2$) in
Fig.~\ref{shear_bed} is comparable to the shear strain rate
applied ($du_x/dy = 0.2$).
When we increased the shear strain rate to $\sim 1.0$, the bed
became homogenized and the region of low particle volume
fraction was not visibly recognizable any more.
Sheared gas fluidized beds are known to fluidize homogeneously
when the shear rate reaches some critical value, depending
on the superficial gas flow rate~\cite{apicella97}.
Our computation yielded comparable values for the critical shear
rate, however, quantitative comparison was not possible, as
the critical rate also depends on the bed height~\cite{apicella97}.

To summarize, we used a Burnett order expression~\cite{sela98} for
normal stress differences (which was developed for the {\it dilute}
limit) to estimate its prediction for {\it dense} suspensions
(as this is the only available theory).
We found that such prediction significantly underestimated
normal stress difference seen in our wave simulations.
Furthermore, when we assessed the first normal stress difference
in beds of perfectly elastic particles, we did not observe any
noticeable decrease (from that in beds of inelastic particles).
In order to computationally obtain the normal stress difference
{\it in sheared dense suspensions} (as opposed to dilute granular
flows in the absence of gas), we considered a sheared fluidized bed.
We found that the normal stress difference between the $xx-$
and $yy-$ components (presumably attributed to Burnett order
effects and/or collisional anisotropy) is still much smaller
than the difference between the $xx-$ and $zz-$ components
which arises from compaction and dilation.
Based on all these results, we conclude that 
($\sigma_{zz}-\sigma_{xx}$) seen in our traveling wave
simulations is not due to Burnett order effects or collisional
anisotropy.

\end{document}